\documentclass[tbtags, onecolumn]{IEEEtran}
\makeatletter
\let\IEEEproof\proof
\let\IEEEendproof\endproof
\let\proof\@undefined
\let\endproof\@undefined
\makeatother
\usepackage[cmex10]{amsmath}
\usepackage{array}
\usepackage{amssymb}
\usepackage{mathrsfs}			
\usepackage{amsthm}
\usepackage{hhline}
\usepackage{threeparttable}
\usepackage{ifthen}
\newtheorem{proposition}{Proposition}
\newtheorem{theorem}{Theorem}
\newcounter{OZC}
\newcommand{\OZN}{\refstepcounter{OZC}{(\alph{OZC})}}
\newcounter{MYtempcnt}
\newcommand{\xn}{\boldsymbol{x}}        
\newcommand{\zn}{\boldsymbol{z}}        
\newcommand{\Sdkr}{\mathscr{S}}		
\newcommand{\Sdklr}{\hat{\Sdkr}}	
\newcommand{\CWS}{A}
\newcommand{\CWSl}{\hat A}
\newcommand{\CCS}{C}
\newcommand{\CCSl}{\hat C}
\newcommand{\chrg}{\sigma}
\newcommand{\rl}[1]{\ifthenelse{\equal{#1}{1}}%
                    {k-d+1}%
                    {(k-d+1)}}          
\newcommand{\At}{\widetilde A}
\newcommand{\Bt}{\widetilde B}
\newcommand{\CCSi}{\CCS(\iota)}
\newcommand{\ats}{p_1}                  
\newcommand{\bts}{p_2}                  
\newcommand{\cts}[1]{\ifthenelse{\equal{#1}{1}}%
                     {\chrg-\ats+\bts}%
                     {(\chrg-\ats+\bts)}} 
\newcommand{\fst}{\mu}                  
\newcommand{\eo}{\rho}			
\newcommand{\fmc}{\left\lfloor-\chrg/2\right\rfloor}
\newcommand{\fpc}{\left\lfloor \chrg/2\right\rfloor}
\newcommand{\F}{{}\,_2^{\vphantom{1}}F_{\!1}^{\vphantom{1}}}
\newcommand{\hyp}[3]{\F\Bigl(
                 \arraycolsep=0em
                 \begin{array}{c}
                  #1\\
                  #2
                 \end{array};
                     #3\Bigr)}
\newcommand{\ufs}{        \left\lceil   \fsr \right\rceil  -1}
\newcommand{\uis}{ \max \left(
                          \left\lceil   \fsr \right\rceil, 0
                        \right)}
\newcommand{\visc}{\max \left(
                         -\left\lceil   \fsr \right\rceil  +1, 1
                        \right)}
\newcommand{\vfsc}{      -\left\lceil   \fsr \right\rceil}
\newcommand{\bd}{b_d}
\newcommand{\bk}{b_k}
\newcommand{\Rt}{\widetilde R}
\DeclareMathOperator*{\Res}{Res}
\newcommand{\rDMB}{\epsilon}         	
\newcommand{\uvc}{\tilde n}
\newcommand{\uvcd}{\tilde n}
\newcommand{\uex}{ \max \left(
                          \uvc, 0
                        \right)}
\newcommand{\vexa}{\max \left(
                         -\uvc, 0
                        \right)}
\newcommand{\pn}{\ensuremath{\boldsymbol{p}}}	
\newcommand{\BRDS}{\vartheta}		
\newcommand{\cl}{c_1}
\newcommand{\cu}{c_2}
\let\proof\IEEEproof
\let\endproof\IEEEendproof
\title{Constant-Weight and Constant-Charge Binary Run-Length Limited Codes}
\author{Oleg~Kurmaev,~\IEEEmembership{Member,~IEEE,}%
 \thanks{The material in Section~\ref{Sec:Recur} of this paper %
         was presented in part at the %
         10th International Workshop on Algebraic %
         and Combinatorial Coding Theory (ACCT-10), %
         Zvenigorod, Russia, September 2006.}
 \thanks{The author is with the Moscow Institute of
         Electronic Engineering (MIEE/MIET),
         124498, Moscow, Russia.
         (e-mail: {\tt \symbol{"3C}kurmaev@org.miet.ru\symbol{"3E}})}}
\date{}
\begin{document}
 \maketitle
\IEEEpeerreviewmaketitle
 \begin{abstract}
  Constant-weight and constant-charge binary sequences
  with constrained run length of zeros are introduced.
  For these sequences, the weight and the charge distribution are found.
  Then, recurrent and direct formulas for calculating the number
  of these sequences are obtained.
  With considering these numbers of constant-weight and constant-charge
  RLL sequences as coefficients of convergent power series,
  generating functions are derived.
  The fact, that generating function
  for enumerating constant-charge RLL sequences
  does not have a closed form, is proved.
  Implementation of encoding and decoding procedures using Cover's
  enumerative scheme is shown.
  On the base of obtained results, some examples,
  such as enumeration of running-digital-sum (RDS) constrained RLL sequences
  or peak-shifts control capability are also provided.
 \end{abstract}


 \section{Introduction}
 The sequences with constrained run length of zeros are known in literature
 as $dk$ sequences.
 In these sequences, single ones are separated by at least $d$,
 but not more then $k$ zeros.
 A $dkr$ sequence is a $dk$ sequence, ending in a run
 of not more then $r$ trailing zeros.
 A $dklr$ sequence is a $dkr$ sequence, beginning with a run of
 not more then $l$ leading zeros.
 These sequences, their properties, and applications
 are described in~\cite{Immink2004}
 in detail or, briefly, in comprehensive overview paper~\cite{ImminkSiegelWolf98}.

 A general enumerative scheme for encoding and decoding binary sequences
 has been presented by Cover~\cite{Cover73}.
 We use this technique for determining the number of constrained sequences.

 Let $\{0, 1\}^n$ be the set of all binary sequences of length $n$
 and let $\xn=(x_1, x_2, \dots, x_n)$ denote a generic element of this set.
 Let $ \Sdkr(n)=\{\xn \in \{0, 1\}^n \; |$
 satisfies the $d, k,    r$ constraints$\}$ and
 let $\Sdklr(n)=\{\xn \in \{0, 1\}^n \; |$
 satisfies the $d, k, l, r$ constraints$\}$.

 Using Cover's method, cardinality of $\Sdklr(n)$
 can be computed as shown in~\cite{Immink97}.
 To do this, the number of $dkr$ sequences,
 which begin with one, is calculated as
 \[
  |\Sdkr(n)|=\sum_{j=d+1}^{k+1} |\Sdkr(n-j)|, \quad n>d+k.
 \]
 Then the number of $dklr$ sequences is calculated as
 \[
  |\Sdklr(n)|=\sum_{j=0}^{\min(n, l)} |\Sdkr(n-j)|.
 \]

 By $\nu=\sum_{j=1}^n x_j$ denote the weight of the sequence $\xn$.
 The number of unconstrained constant-weight sequences
 may be simply obtained as $\binom n \nu$, see~\cite{Schalkwijk72}.
 Methods for calculating the number of constant-weight $dkr$ sequences
 is given by~\cite{Ytrehus91, Kurmaev02en}.

 Under NRZI encoding~\cite{ImminkSiegelWolf98}
 we understand mapping the source sequence $\xn$
 to bipolar sequence $\zn$, $\zn \in \{-1, 1\}^n$
 such that
 \begin{align*}
  z_j &=\begin{cases}
    z_{j-1}, & x_j=0,\\
   -z_{j-1}, & x_j=1,
  \end{cases}\\
  z_0 &=1.
 \end{align*}

 By $\chrg=\sum_{j=1}^n z_j$ denote the digital sum or charge
 of the sequence $\zn$. Observe that $\chrg \in [-n, n]$, where $\chrg$
 admits even values whenever $n$ is even and odd values whenever $n$ is odd.
 For example, Table~\ref{Tab:Alldklr} shows sequences $\xn$ and $\zn$,
 theirs $\nu$ and $\chrg$
 respectively.

 \begin{table}
   \centering
  \begin{threeparttable}
   \caption{All Lexicographically Ordered $dklr$ Sequences
    of Length $n=8$, With Constraints $d=2$, $k=4$, $l=1$, $r=3$.}
   \label{Tab:Alldklr}
   \tabcolsep=0.5em
   \renewcommand{\arraystretch}{0}
   \newcommand{\z}{$\scriptstyle1$}
   \newcommand{\xs}{\rule{0pt}{2ex}}
   \newcommand{\zs}{\rule{0pt}{1ex}}
   \newcommand{\zf}{\rule{0pt}{2pt}}
   \begin{tabular}{|l|cccccccc|lr|}
    \hline
    \strut %
    $N\tnote{a}$ %
        & $x_1$ & $x_2$ & $x_3$ & $x_4$ & %
          $x_5$ & $x_6$ & $x_7$ & $x_8$ & $\nu$ &        \\
        & $z_1$ & $z_2$ & $z_3$ & $z_4$ & %
          $z_5$ & $z_6$ & $z_7$ & $z_8$ & \strut& $\chrg$\\
    \hhline{|=|========|==|}
    \xs%
    0  &  0&  1&  0&  0&  0&  0&  1&  0 &   2   &    \\
    \zs%
       & \z&-\z&-\z&-\z&-\z&-\z& \z& \z &       &  -2\\
    \zf&   &   &   &   &   &   &   &    &       &    \\
    \hline
    \xs%
    1  &  0&  1&  0&  0&  0&  1&  0&  0 &   2   &    \\
    \zs%
       & \z&-\z&-\z&-\z&-\z& \z& \z& \z &       &   0\\
    \zf&   &   &   &   &   &   &   &    &       &    \\
    \hline
    \xs%
    2  &  0&  1&  0&  0&  1&  0&  0&  0 &   2   &    \\
    \zs%
       & \z&-\z&-\z&-\z& \z& \z& \z& \z &       &   2\\
    \zf&   &   &   &   &   &   &   &    &       &    \\
    \hline
    \xs%
    3  &  0&  1&  0&  0&  1&  0&  0&  1 &   3   &    \\
    \zs%
       & \z&-\z&-\z&-\z& \z& \z& \z&-\z &       &   0\\
    \zf&   &   &   &   &   &   &   &    &       &    \\
    \hline
    \xs%
    4  &  1&  0&  0&  0&  0&  1&  0&  0 &   2   &    \\
    \zs%
       &-\z&-\z&-\z&-\z&-\z& \z& \z& \z &       &  -2\\
    \zf&   &   &   &   &   &   &   &    &       &    \\
    \hline
    \xs%
    5  &  1&  0&  0&  0&  1&  0&  0&  0 &   2   &    \\
    \zs%
       &-\z&-\z&-\z&-\z& \z& \z& \z& \z &       &   0\\
    \zf&   &   &   &   &   &   &   &    &       &    \\
    \hline
    \xs%
    6  &  1&  0&  0&  0&  1&  0&  0&  1 &   3   &    \\
    \zs%
       &-\z&-\z&-\z&-\z& \z& \z& \z&-\z &       &  -2\\
    \zf&   &   &   &   &   &   &   &    &       &    \\
    \hline
    \xs%
    7  &  1&  0&  0&  1&  0&  0&  0&  1 &   3   &    \\
    \zs%
       &-\z&-\z&-\z& \z& \z& \z& \z&-\z &       &   0\\
    \zf&   &   &   &   &   &   &   &    &       &    \\
    \hline
    \xs%
    8  &  1&  0&  0&  1&  0&  0&  1&  0 &   3   &    \\
    \zs%
       &-\z&-\z&-\z& \z& \z& \z&-\z&-\z &       &  -2\\
    \zf&   &   &   &   &   &   &   &    &       &    \\
    \hline
   \end{tabular}
   \begin{tablenotes}
    \item [a] By $N$ we denote the lexicographic index of sequence.
   \end{tablenotes}
  \end{threeparttable}
 \end{table}

 Our central goal will be to determine the number of
 constant-weight and constant-charge run length limited sequences
 as well as an exact estimation of this number.
 We also intend to show that using Cover's enumerative method
 provides us with necessary values
 for error control.
 Namely, these values follow from weight distribution,
 run-length distribution, and charge distribution,
 which may be obtained by Cover's technique.

 Although we will consider constant-weight
 and constant-charge RLL sequences together,
 our main efforts will be focused on constant-charge RLL sequences.
 Some results concerning the constant-weight RLL sequences
 are known with the contributions coming from
 Lee~\cite{Lee88}, Forsberg and Blake~\cite{Forsberg88}, Ytrehus~\cite{Ytrehus91}.
 We will cite some of their results for the sake of generalization
 and in comparison with the similar results for constant-charge codes.

 Run-length distribution does not considered in this paper,
 although the length of each run is accounted in Cover's enumerative technique.
 The problems of defining this distribution and accompanying problems
 are studied in source coding.
 The reader can find these materials in various publications
 since well-known Huffman's paper~\cite{Huffman52}.

 The rest of this paper will be organized as follows.
 First, in Section~\ref{Sec:Recur},
 we derive recursion relations
 for determining the number of the sequences.
 Further, in Section~\ref{Sec:Direct},
 we obtain direct formulas for the same.
 Next, in Section~\ref{Sec:GenFunct},
 we consider the numbers of our sequences
 as coefficients of formal power series and derive
 generating functions.
 We also prove that generating function
 for enumerating constant-charge RLL
 sequences does not have a closed form.
 Then, in Section~\ref{Sec:EnumAlg},
 we give an enumerative algorithm for encoding
 and decoding these sequences.
 Some remarks and application notes are presented in Section~\ref{Sec:FuRem}.
 Finally, conclusions are drawn in Section~\ref{Sec:Concl}.
%
%

 \section{The Number of Sequences} \label{Sec:Recur}

 Consider run-length constrained binary sequences
 of length $n$ and weight $\nu$.
 Let $\CWS_n^\nu$ be the number of these sequences
 which begin with one.
 Let $\CWSl_n^\nu$ be the number of these sequences
 which begin with a leading run of zeros.

 Since an internal run of zeros succeeds a leading run of zeros
 then the leading constraint $l$ does not affect $\CWS_n^\nu$.
 For convenience, below under $\CWS_n^\nu$
 we consider $\CWS_n^\nu (d, k, r)$
 and under $\CWSl_n^\nu$
 we similarly consider $\CWSl_n^\nu (d, k, l, r)$.

 Suppose that a unique sequence of zero length and zero weight exists.
 Let it be a sequence which begins with one. Then
 \[
  \CWS_0^0=1.
 \]

 \begin{proposition}
  \label{Prop:A}
  The numbers $\CWS_n^\nu$ and $\CWSl_n^\nu$
  can be obtained as:\\
  If $\nu=1$ and the sequences begin with one, then
  \begin{equation}
   \label{Eq:A1}
   \CWS_n^1=\begin{cases}
             1, & 1\leq n \leq r+1,\\
             0, & \text{otherwise}.
            \end{cases}
  \end{equation}
  If $\nu>1$ and the sequences begin with one, then
  \begin{equation}
   \label{Eq:Anu}
   \CWS_n^\nu=\begin{cases}
               0,                                              & n<d+1,\\
               \sum_{j=d+1}^{\min(n, k+1)} \CWS_{n-j}^{\nu-1}, & d+1\leq n.
              \end{cases}
  \end{equation}
  If $\nu=0$ and a leading series of zeros is running, then
  \begin{equation}
   \label{Eq:AL0}
   \CWSl_n^0=\begin{cases}
              1, & n\leq\min(l, r),\\
              0, & \text{otherwise}.
             \end{cases}
  \end{equation}
  If $\nu>0$ and a leading series of zeros is running, then
  \begin{equation}
   \label{Eq:ALnu}
   \CWSl_n^\nu (l, r)=\sum_{j=0}^{\min(n, l)} \CWS_{n-j}^\nu.
  \end{equation}
 \end{proposition}
 Here $d\geq 0$, $k\geq d$, $l\geq 0$, $r\geq 0$.
 \begin{IEEEproof}
  In the case of $\nu=1$,
  there is only a trailing run of zeros in the sequence.
  It gives us the only allowed sequence
  which length lies in the interval $[1, r+1]$.
  Therefore, \eqref{Eq:A1} is evident.
  In the case of $\nu>1$,
  according to Cover's enumerative method~\cite{Cover73},
  we build the recursion by the following way.
  Let us consider a possible run of zeros, which follows the leading one,
  as a prefix for the following subsequences beginning also with one.
  Assuming the length of the prefix grows from $d+1$ to $\min(n, k+1)$
  and weight of this prefix equals one, we obtain
  the number of these subsequences evidently equal $\CWS_{n-j}^{\nu-1}$
  for each allowed prefix and the total number is expressed by~\eqref{Eq:Anu}.

  The other case is when a leading series of zeros is running.
  In the case of zero weight, the leading run of zeros is the trailing one.
  This also gives us the only allowed sequence
  which length lies in the interval $[0, \min(l, r)]$.
  Hence, \eqref{Eq:AL0} is also evident.
  In the case of nonzero weight, there exist only zero weight prefixes
  which length lies in the interval $[0, \min(n, l)]$.
  Subsequences beginning with one follow this prefixes,
  thus, the total number of sequences is expressed by~\eqref{Eq:ALnu}.
 \end{IEEEproof}

 Consider bipolar run-length constrained sequences
 of length $n$ and charge $\chrg$.
 We can do this in terms of source sequence $\xn$.
 This allows us to obtain results similar to Proposition~\ref{Prop:A}.
 In this case $\chrg=\sum_{j=1}^n(-1)^{\nu_j}$,
 where $\nu_j=\sum_{i=1}^jx_i$.
 Let $\CCS_n^\chrg$ be the number of these sequences,
 which begin with one.
 Let $\CCSl_n^\chrg$ be the number of these sequences,
 which begin with a leading run of zeros.

 Since an internal run of zeros succeeds a leading run of zeros
 then the leading constraint $l$ does not affect $\CCS_n^\chrg$.
 For convenience, below under $\CCS_n^\chrg$
 we consider $\CCS_n^\chrg (d, k, r)$
 and under $\CCSl_n^\chrg$
 we similarly consider $\CCSl_n^\chrg (d, k, l, r)$.

 \begin{proposition}
  \label{Prop:C}
  The numbers $\CCS_n^\chrg$ and $\CCSl_n^\chrg$
  can be obtained as:\\
  If $\chrg=-n$ and the sequences begin with one, then
  \begin{equation}
   \label{Eq:Cmn}
   \CCS_n^{-n}=\begin{cases}
                1, & n\leq r+1,\\
                0, & \text{otherwise}.
               \end{cases}
  \end{equation}
  If $\chrg \neq -n$ and the sequences begin with one, then
  \begin{equation}
   \label{Eq:Cc}
   \CCS_n^\chrg=\begin{cases}
                 0,                          & n<d+1,\\
                 \sum_{j=d+1}^{\min(n, k+1)}
                  \CCS_{n-j}^{-\chrg-j},     & d+1\leq n.
                \end{cases}
  \end{equation}
  If $\chrg=n$ and a leading series is running, then
  \[
   \CCSl_n^n=\begin{cases}
              1, & n\leq\min(l, r),\\
              0, & \text{otherwise}.
             \end{cases}
  \]
  If $\chrg \neq n$ and a leading series is running, then
  \begin{equation}
   \label{Eq:Cn}
   \CCSl_n^\chrg=\sum_{j=0}^{\min(n, l)} \CCS_{n-j}^{\chrg-j}.
  \end{equation}
 \end{proposition}
 Here $d\geq 0$, $k\geq d$, $l\geq 0$, $r\geq 0$.
 \begin{IEEEproof}
  Is similar to Proposition~\ref{Prop:A},
  except the differences between the weight and charge.
  Namely, if the sequences of length $n$ begin with one,
  then the charge $\chrg_n$ can be obtained as
  \[
   \chrg_n=-n+m-\chrg_m,
  \]
  where $m$ and $\chrg_m$ is the length and charge of following subsequence
  that also begins with one. Since $m=n-j$,
  therefore, the number of subsequences in~\eqref{Eq:Cc}
  must be $\CCS_{n-j}^{-\chrg-j}$.

  If the sequences of length $n$ begin with zero,
  then the charge $\chrg_n$ can be obtained as
  \[
   \chrg_n=n-m+\chrg_m.
  \]
  Thus, the number of subsequences in~\eqref{Eq:Cn}
  must be $\CCS_{n-j}^{\chrg-j}$.
 \end{IEEEproof}
 For example, the charge changing is shown in Table~\ref{Tab:ChrgChng}.

 \begin{table}[!t]
  \renewcommand{\arraystretch}{1.3}
  \caption{An Example of Charge Changing}
  \label{Tab:ChrgChng}
  \centering
  \newcommand{\z}{$\scriptstyle1$}
  \tabcolsep=0.25em
  \begin{tabular}{c|r|cccccccc|c|}
   \multicolumn{11}{l}{\OZN~After a leading one.}     \\
   \cline{2-11}
   & $m  $ & $   $ & $   $ & $   $ & $x_1$ & %
             $x_2$ & $x_3$ & $x_4$ & $x_5$ & $\chrg_m$\\
   \cline{2-11}
   & 5  &   &   &   &  1&  0&  0&  1&  0       &      \\
   &    &   &   &   &-\z&-\z&-\z& \z& \z       &  -1  \\
   \hhline{~|=|========|=|}
   & $n  $ & $x_1$ & $x_2$ & $x_3$ & $x_4$ & %
             $x_5$ & $x_6$ & $x_7$ & $x_8$ & $\chrg_n$\\
   \cline{2-11}
   & 8  &  1&  0&  0&  1&  0&  0&  1&  0       &      \\
   &    &-\z&-\z&-\z& \z& \z& \z&-\z&-\z       &  -2  \\
   \cline{2-11}
  \end{tabular}
  \begin{tabular}{c|r|cccccccc|c|}
   \multicolumn{11}{l}{\OZN~After leading zeros.}     \\
   \cline{2-11}
   & $m  $ & $   $ & $   $ & $   $ & $x_1$ & %
             $x_2$ & $x_3$ & $x_4$ & $x_5$ & $\chrg_m$\\
   \cline{2-11}
   & 5  &   &   &   &  1&  0&  0&  1&  0       &      \\
   &    &   &   &   &-\z&-\z&-\z& \z& \z       &  -1  \\
   \hhline{~|=|========|=|}
   & $n  $ & $x_1$ & $x_2$ & $x_3$ & $x_4$ & %
             $x_5$ & $x_6$ & $x_7$ & $x_8$ & $\chrg_n$\\
   \cline{2-11}
   & 8  &  0&  0&  0&  1&  0&  0&  1&  0       &      \\
   &    & \z& \z& \z&-\z&-\z&-\z& \z& \z       &   2  \\
   \cline{2-11}
  \end{tabular}
  \setcounter{OZC}{0}
 \end{table}

 Using relations derived in this section,
 we can write the weight and charge distribution of our sequences.
 It seems to us that mostly convenient form
 for presenting this distribution is a triangle table,
 like Pascal's triangle.
 An example of such distribution is shown in Table~\ref{Tab:ChrgDistr}.

 \begin{table}[!t]
  \renewcommand{\arraystretch}{1.3}
  \caption{An Example of Weight and Charge Distribution}
  \label{Tab:ChrgDistr}
  \centering
  \tabcolsep=0.2em
  \begin{tabular}{c|l|ccccccccc@{$\;$}|ccccccccccccccccc|}
   \multicolumn{28}{l}{\OZN~For sequences
    beginning with one.} \setcounter{MYtempcnt}{\value{OZC}} \\
   \multicolumn{28}{l}{\hspace{1em}
    Constraints:~$d=2$, $k=4$, $r=3$.} \\
   \cline{2-28}
   & \multicolumn{1}{c|}{\vphantom{$\hat X^X$}} & %
     \multicolumn{9}{c|}{$\CWS_n^\nu$} & %
     \multicolumn{17}{c|}{$\CCS_n^\chrg$} \\
   \cline{2-28}
   & \multicolumn{1}{r|}{\quad $\nu, \chrg$} & %
          0&1&2&3&4&5&6&7&8&  -8&\!-7&\!-6&\!-5&\!-4&\!-3&\!-2&\!%
                              -1&\,0&\,1&\,2&\,3&\,4&\,5&\,6&\,7&\,8\\
   & $n$&  & & & & & & & & &    & & & & & & & & & & & & & & & & \\
   \hhline{~|=|=========|=================|}
   & 0  & 1& & & & & & & & &    & & & & & & & &1& & & & & & & & \\
   & 1  & 0&1& & & & & & & &    & & & & & & &1& &0& & & & & & & \\
   & 2  & 0&1&0& & & & & & &    & & & & & &1& &0& &0& & & & & & \\
   & 3  & 0&1&0&0& & & & & &    & & & & &1& &0& &0& &0& & & & & \\
   & 4  & 0&1&1&0&0& & & & &    & & & &1& &1& &0& &0& &0& & & & \\
   & 5  & 0&0&2&0&0&0& & & &    & & &0& &1& &1& &0& &0& &0& & & \\
   & 6  & 0&0&3&0&0&0&0& & &    & &0& &1& &1& &1& &0& &0& &0& & \\
   & 7  & 0&0&3&1&0&0&0&0& &    &0& &0& &1& &2& &1& &0& &0& &0& \\
   & 8  & 0&0&2&3&0&0&0&0&0& \,0& &0& &0& &3& &2& &0& &0& &0& &0\\
   \cline{2-28}
  \end{tabular}
  \begin{tabular}{c|l|ccccccccc@{$\;$}|ccccccccccccccccc|}
   \multicolumn{28}{l}{\OZN~For sequences
    beginning with a leading run of zeros.} \\
   \multicolumn{28}{l}{\hspace{1em}
    Constraints:~$d=2$, $k=4$, $l=1$, $r=3$.} \\
   \cline{2-28}
   & \multicolumn{1}{c|}{} & %
     \multicolumn{9}{c|}{$\CWSl_n^\nu$} & %
     \multicolumn{17}{c|}{$\CCSl_n^\chrg$} \\
   \cline{2-28}
   & \multicolumn{1}{r|}{\quad $\nu, \chrg$} & %
          0&1&2&3&4&5&6&7&8&  -8&\!-7&\!-6&\!-5&\!-4&\!-3&\!-2&\!%
                              -1&\,0&\,1&\,2&\,3&\,4&\,5&\,6&\,7&\,8\\
   & $n$&  & & & & & & & & &    & & & & & & & & & & & & & & & & \\
   \hhline{~|=|=========|=================|}
   & 0  & 1& & & & & & & & &    & & & & & & & &1& & & & & & & & \\
   & 1  & 1&1& & & & & & & &    & & & & & & &1& &1& & & & & & & \\
   & 2  & 0&2&0& & & & & & &    & & & & & &1& &1& &0& & & & & & \\
   & 3  & 0&2&0&0& & & & & &    & & & & &1& &1& &0& &0& & & & & \\
   & 4  & 0&2&1&0&0& & & & &    & & & &1& &2& &0& &0& &0& & & & \\
   & 5  & 0&1&3&0&0&0& & & &    & & &0& &2& &2& &0& &0& &0& & & \\
   & 6  & 0&0&5&0&0&0&0& & &    & &0& &1& &2& &2& &0& &0& &0& & \\
   & 7  & 0&0&6&1&0&0&0&0& &    &0& &0& &2& &3& &2& &0& &0& &0& \\
   & 8  & 0&0&5&4&0&0&0&0&0& \,0& &0& &0& &4& &4& &1& &0& &0& &0\\
   \cline{2-28}
  \end{tabular}
  \setcounter{OZC}{0}
  \vspace*{-8pt}
 \end{table}

 We may consider~\eqref{Eq:Cc} as an implicit mutual recursion.
 Indeed, an element in left slanting row in the triangle of charge
 distribution (see Table~\ref{Tab:ChrgDistr}(\alph{MYtempcnt}))
 depends on elements in right slanting rows and vice versa.
 In more details, this will be shown in the next section.

 \section{Direct Equations for the Number of Sequences} \label{Sec:Direct}

 Consider $dkr$~- limited sequences.
 The problem of defining the number of constant-weight sequences,
 in which term's coefficients are Fibonacci numbers,
 i.e., for $d$~- sequences, has been solved by Riordan~\cite{Riordan58}.
 He presented a direct, not recursion method for calculating the number
 of constant-weight $d$~- sequences.
 Lee in~\cite{Lee88} and Ytrehus in~\cite{Ytrehus91}
 extended Riordan's method for the $dkr$~- constrained sequences.
 Our method of deriving the similar direct equations rests on recursions
 from the previous section.

 \subsection{Calculating the number of constant-weight sequences}

 From~\eqref{Eq:A1} we have that
 \[
  \CWS_1^1=1, \CWS_2^1=1, \dots, \CWS_{r+1}^1=1, \CWS_{r+2}^1=0,\dots
 \]
 or
 \[
  \CWS_n^1=\binom{n-1}0-\binom{n-1-(r+1)}0
 \]
 (suppose that $\binom nk=0$ if $k>n$).
 From recursion relation~\eqref{Eq:Anu} we have
 \begin{align*}
  \CWS_n^2   & =\CWS_{n-(d+1)}^1      +\CWS_{n-(d+2)}^1      +\dots %
               +\CWS_{n-(k+1)}^1,\\
  \CWS_n^3   & =\CWS_{n-(d+1)}^2      +\CWS_{n-(d+2)}^2      +\dots %
               +\CWS_{n-(k+1)}^2,\\
  \vdots\;\; & \quad\;\;\:\vdots\hphantom{%
               =\CWS_{n-(d+1)}^1}\vdots\hphantom{%
                                      +\CWS_{n-(d+2)}^1      +\dots}\vdots\\
  \CWS_n^\nu & =\CWS_{n-(d+1)}^{\nu-1}+\CWS_{n-(d+2)}^{\nu-1}+\dots %
               +\CWS_{n-(k+1)}^{\nu-1}.
 \end{align*}
 Substituting $\CWS_n^1$ into $\CWS_n^2$, we obtain
 \begin{align}
  \begin{split}
   \CWS_n^2    &  =     \tbinom{n-(d+1)-1}0-\tbinom{n-(d+1)-1-(r+1)}0 \\
               & \qquad%
                     {}+\tbinom{n-(d+2)-1}0-\tbinom{n-(d+2)-1-(r+1)}0 \\
               & \qquad%
                 \cdots+\tbinom{n-(k+1)-1}0-\tbinom{n-(k+1)-1-(r+1)}0 \\
               &  =       \tbinom{n        -(d+1)}1
                         -\tbinom{n  -(r+1)-(d+1)}1               \\
               & \qquad%
                 {}-\left(\tbinom{n-1      -(k+1)}1
                         -\tbinom{n-1-(r+1)-(k+1)}1\right).
  \end{split}
                                                                \notag\\
  \intertext{Likewise,}
  \begin{split}
   \CWS_n^3    &  =       \tbinom{n+1      -2(d+1)       }2
                         -\tbinom{n+1-(r+1)-2(d+1)       }2       \\
               & \qquad%
                {}-2\left(\tbinom{n         -(d+1) -(k+1)}2
                         -\tbinom{n  -(r+1) -(d+1) -(k+1)}2\right)\\
               & \qquad%
                       {}+\tbinom{n-1             -2(k+1)}2
                         -\tbinom{n-1-(r+1)       -2(k+1)}2,
  \end{split}
                                                                \notag\\
  \vdots\;\;\; & \mspace{32mu}  \vdots%
                 \mspace{72mu}  \vdots%
                 \mspace{120mu} \vdots \notag\\
  \begin{split}
   \label{Eq:AdkrOrig}
   \CWS_n^\nu  &  =\sum_{j=0}^{\nu-1}(-1)^j \tbinom{\nu-1}j
                    \left(\tbinom{n-1+(\nu-1-j)%
                                   -(\nu-1-j)(d+1)-j(k+1)}{\nu-1} \right. \\
               &     \qquad                                       \left.
                       {}-\tbinom{n-1+(\nu-1-j)-(r+1)%
                                   -(\nu-1-j)(d+1)-j(k+1)}{\nu-1} \right),\\
               & \mspace{320mu}
                     \quad \nu \ge 1.
  \end{split}
 \end{align}
 By $ q $ denote the number of possible lengths of runs beginning with one
 \renewcommand{\rl}[1]{q}              
 \begin{equation*}
  \rl1=k-d+1.
 \end{equation*}
 Then we can rewrite~\eqref{Eq:AdkrOrig} in more convenient form
 \begin{equation}
  \label{Eq:Adkr}
  \begin{split}
   \CWS_n^\nu &= \sum_{j=0}^{\nu-1}(-1)^j \tbinom{\nu-1}j
                  \left(\tbinom{n-1      -(\nu-1)d-j\rl2}{\nu-1} \right. \\
              &    \qquad \qquad                                 \left.
                     {}-\tbinom{n-1-(r+1)-(\nu-1)d-j\rl2}{\nu-1} \right), %
                   \quad \nu \ge 1.
  \end{split}
 \end{equation}

 \subsection{Calculating the number of constant-charge sequences}

 Now, consider the constant-charge sequences. For these sequences we may simply
 show that
 \[
  \CCS_0^0=1, \CCS_1^{-1}=1, \dots, \CCS_{r+1}^{-(r+1)}=1,
  \CCS_{r+2}^{-(r+2)}=0,     \dots
 \]
 or
 \begin{equation}
  \label{Eq:Clsr_dkr}
  \CCS_n^{-n}=\binom n0-\binom{n-1-(r+1)}0.
 \end{equation}
 This directly follows from initial conditions~\eqref{Eq:Cmn}
 of recursion relation~\eqref{Eq:Cc}.

 Thus, we define the first left slanting row in our triangle of numbers
 of constant-charge sequences.
 For example, see Table~\ref{Tab:ChrgDistr}(\alph{MYtempcnt}).

 We also may show that for right slanting rows
 \begin{equation}
  \label{Eq:Crsr_n_n_dkr}
  \begin{alignedat}{3}
   \CCS_0^0          & =1, \; & \CCS_1^1              & =0, %
    \; & \CCS_2^2              & =0, %
     \dots \quad d \geq 0,   \\
   \CCS_1^{-1}       & =1, \; & \CCS_2^0              & =0, %
    \; & \CCS_3^1              & =0, %
     \dots \quad d \geq 1,   \\
   \vdots\quad       &        & \vdots\;\;            &     %
       & \vdots\;\;            &     %
     \qquad\qquad\;\;\vdots \\
   \CCS_{r+1}^{-r+1} & =1, \; & \CCS_{r+2}^{-(r+1)+1} & =0, %
    \; & \CCS_{r+3}^{-(r+1)+2} & =0, %
     \dots \quad d \geq r+1, \\
   \CCS_{r+2}^{-r+2} & =0, \; & \CCS_{r+3}^{-(r+2)+1} & =0, %
    \; & \CCS_{r+4}^{-(r+2)+2} & =0, %
     \dots
  \end{alignedat}
 \end{equation}

 Let $\delta$ be an even number which shows how the charge $\chrg$
 differs from $n$ or from $-n$.
 In other words, $\delta/2$ is an index of a slanting row in the triangle
 of numbers of constant-charge sequences.

 Therefore, we can rewrite~\eqref{Eq:Clsr_dkr} as
 \begin{equation}
  \label{Eq:Clsr_dkr1}
  \begin{split}
   \CCS_n^{-n+\delta} &= \left(\binom{\delta/2}0-\binom{\delta/2-1}0\right) \\
                      &  \qquad \qquad
                        {}\times%
                           \left(\binom n0-\binom{n-1-(r+1)}0\right).
  \end{split}
 \end{equation}
 Also we can rewrite~\eqref{Eq:Crsr_n_n_dkr} as
 \begin{equation}
  \label{Eq:Crsr_dkr}
  \begin{split}
   \CCS_n^{n-\delta} &= \left(\binom{\delta/2-1}0-\binom{\delta/2-1-(r+1)}0\right) \\
                     &  \qquad
                       {}\times%
                          \left(\binom{n-\delta/2}0-\binom{n-\delta/2-1}0\right),
                           \quad
                          d \geq \delta/2.
  \end{split}
 \end{equation}
 If $d < \delta/2$ then for right slanting rows,
 we have from recursion relation~\eqref{Eq:Cc} that
 \begin{align}
  \label{Eq:Crsrs_dkr}
   \CCS_n^{n-\delta} & =\CCS_{n-(d+1)}^{-n+\delta-(d+1)}%
                       +\CCS_{n-(d+2)}^{-n+\delta-(d+2)}%
                       +\dots%
                       +\CCS_{n-(k+1)}^{-n+\delta-(k+1)}          \notag \\
                     & =\CCS_{n-(d+1)}^{-(n-(d+1))+\delta-2(d+1)}%
                       +\CCS_{n-(d+2)}^{-(n-(d+2))+\delta-2(d+2)} \notag \\
                     &  \cdots%
                       +\CCS_{n-(k+1)}^{-(n-(k+1))+\delta-2(k+1)}.
 \end{align}
 This confirms a mutual nature of~\eqref{Eq:Cc}.
 The series~\eqref{Eq:Crsrs_dkr} terminates early if $\chrg < -n$,
 i.e., whenever someone of $\delta-2(d+1)$, $\delta-2(d+2)$,~\dots, $\delta-2(k+1)$
 becomes less then $0$.
 Therefore, combining~\eqref{Eq:Crsr_dkr} with~\eqref{Eq:Crsrs_dkr}, we obtain
 \begin{equation*}
  \begin{split}
   \CCS_n^{n-\delta} &   =\CCS_{n-(d+1)}^{-(n-(d+1))+\delta-2(d+1)}%
                         +\CCS_{n-(d+2)}^{-(n-(d+2))+\delta-2(d+2)}                \\
                     &    \cdots%
                         +\CCS_{n-(k+1)}^{-(n-(k+1))+\delta-2(k+1)}                \\
                     & {}+\left(\tbinom{\delta/2-1}0
                               -\tbinom{\delta/2-1-(r+1)}0\right)
                          \left(\tbinom{n-\delta/2}0-\tbinom{n-\delta/2-1}0\right) \\
                     & {}-\left(\tbinom{\delta/2  -(d+1)}0
                               -\tbinom{\delta/2-1-(d+1)}0\right.                  \\
                       \vphantom{%
                       {}-\left(\tbinom{\delta/2  -(d+1)}0
                               -\tbinom{\delta/2-1-(d+1)}0\right.}
                     &       {}+\tbinom{\delta/2  -(d+2)}0
                               -\tbinom{\delta/2-1-(d+2)}0                         \\
                     &    \left.
                         \cdots+\tbinom{\delta/2  -(k+1)}0
                               -\tbinom{\delta/2-1-(k+1)}0\right)                  \\
                     & {}\times%
                          \left(\tbinom{n-\delta/2}0-\tbinom{n-\delta/2-1}0\right),
                                                      \quad \delta > 0,
  \end{split}
 \end{equation*}
 then
 \begin{equation}
  \label{Eq:Crsrs_dkr1}
  \begin{split}
   \CCS_n^{n-\delta} &   =\CCS_{n-(d+1)}^{-(n-(d+1))+\delta-2(d+1)}%
                         +\CCS_{n-(d+2)}^{-(n-(d+2))+\delta-2(d+2)}    \\
                     &    \cdots%
                         +\CCS_{n-(k+1)}^{-(n-(k+1))+\delta-2(k+1)}    \\
                     & {}+\left(\tbinom{\delta/2-1}0
                               -\tbinom{\delta/2-(d+1)}0\right.        \\
                     &    \left.{}%
                               +\tbinom{\delta/2-1-(k+1)}0
                               -\tbinom{\delta/2-1-(r+1)}0\right)      \\
                     & {}\times%
                          \left(\tbinom{n-\delta/2}0-\tbinom{n-\delta/2-1}0\right),
                                                      \quad \delta > 0.
  \end{split}
 \end{equation}

 Consider an example.
 For $\delta=2$ and for $\chrg=n-\delta$,
 we have from recursion relation~\eqref{Eq:Cc} that
 \begin{equation*}
  \begin{split}
   \CCS_n^{n-2} & =\CCS_{n-(d+1)}^{-n+2-(d+1)}%
                  +\CCS_{n-(d+2)}^{-n+2-(d+2)}%
                  +\dots%
                  +\CCS_{n-(k+1)}^{-n+2-(k+1)}\\
                & =\CCS_{n-(d+1)}^{-(n-(d+1))+2-2(d+1)}%
                  +\CCS_{n-(d+2)}^{-(n-(d+2))+2-2(d+2)}\\
                &  \cdots%
                  +\CCS_{n-(k+1)}^{-(n-(k+1))+2-2(k+1)}
  \end{split}
 \end{equation*}
 or
 \[
  \CCS_n^{n-2}=\CCS_{n-1}^{-(n-1)}, \quad d=0.
 \]
 Since this equation specifies the right slanting rows,
 then we have to substitute~\eqref{Eq:Clsr_dkr} for $\CCS_{n-1}^{-(n-1)}$, if $d=0$,
 and~\eqref{Eq:Crsr_dkr} otherwise.
 Therefore we obtain the next result
 \[
  \CCS_n^{n-2}=\begin{cases}
   \binom{n-1}0-\binom{n-1-1-(r+1)}0, & d=0,\\
    \binom{n-1}0-\binom{n-1-1}0,      & \text{otherwise}.
  \end{cases}
 \]

 Using~\eqref{Eq:Crsrs_dkr1}, we can rewrite this piecewise-continuous equation as
 \begin{equation*}
  \begin{split}
   & \begin{split}
      \CCS_n^{n-2} &=  \left(      \tbinom{1+1-(d+1)}1-\tbinom{1  -(d+1)}1
                       \right.                                                    \\
                   &   \qquad
                       \left.{}%
                            -\left(\tbinom{1  -(k+1)}1-\tbinom{1-1-(k+1)}1\right)
                       \right)                                                    \\
                   &   \qquad \quad
                     {}\times%
                       \left(\tbinom{n-1}0-\tbinom{n-1-1-(r+1)}0\right)
     \end{split}                                                                  \\[2ex]
   & \quad
     \begin{split}
                   &{}+\left(\tbinom{1-1}0
                            -\tbinom{1-(d+1)}0
                            +\tbinom{1-1-(k+1)}0
                            -\tbinom{1-1-(r+1)}0\right)                           \\
                   &   \qquad
                     {}\times%
                       \left(\tbinom{n-1}0-\tbinom{n-1-1\vphantom{()}}0\right).
     \end{split}
  \end{split}
 \end{equation*}
 For $\delta=2$ and for $\chrg=-n+\delta$,
 we have from the same recursion relation
 \begin{equation*}
  \CCS_n^{-n+2}=\CCS_{n-(d+1)}^{n-(d+1)-2}%
               +\CCS_{n-(d+2)}^{n-(d+2)-2}%
               +\dots%
               +\CCS_{n-(k+1)}^{n-(k+1)-2}
 \end{equation*}
 and obtain
 \begin{equation*}
  \begin{split}
   & \begin{split}
      \CCS_n^{-n+2} &=  \left(      \tbinom{1+1-(d+1)}1
                                   -\tbinom{1  -(d+1)}1       \right.         \\
                    &   \qquad
                        \left.{}%
                             -\left(\tbinom{1  -(k+1)}1
                                   -\tbinom{1-1-(k+1)}1\right)\right)         \\
                    &
                      {}\times%
                        \left(      \tbinom{n-1+1      -(d+1)}1
                                   -\tbinom{n-1  -(r+1)-(d+1)}1       \right. \\
                    &   \qquad
                                                                      \left.
                           {}-\left(\tbinom{n-1        -(k+1)}1
                                   -\tbinom{n-1-1-(r+1)-(k+1)}1\right)\right)
     \end{split}                                                              \\[2ex]
   & \quad
     \begin{split}
                    &{}+\left(\tbinom{1-1}0
                             -\tbinom{1-(d+1)}0
                             +\tbinom{1-1-(k+1)}0
                             -\tbinom{1-1-(r+1)}0\right)                      \\
                    &   \qquad
                      {}\times%
                        \left(      \tbinom{n-1+1-(d+1)}1
                                   -\tbinom{n-1  -(d+1)}1       \right.       \\
                    &   \qquad \quad
                        \left.{}%
                             -\left(\tbinom{n-1  -(k+1)}1
                                   -\tbinom{n-1-1-(k+1)}1\right)\right).
     \end{split}
  \end{split}
 \end{equation*}

 Let $ A $ be a value which has the similar combinatorial meaning
 as $ A $ in~\eqref{Eq:Anu} and let
 \begin{equation}
  \label{Eq:A}
  \begin{split}
   A_\delta^m &= \sum_{j=0}^m (-1)^j \tbinom mj
                      \left(\tbinom{\delta/2+m-j  -(m-j)(d+1)-j(k+1)}m
                       \vphantom{%
                           -\tbinom{\delta/2+m-j-1-(m-j)(d+1)-j(k+1)}m}\right. \\
              &       \qquad
                                                                       \left.
                       \vphantom{%
                            \tbinom{\delta/2+m-j  -(m-j)(d+1)-j(k+1)}m}
                         {}-\tbinom{\delta/2+m-j-1-(m-j)(d+1)-j(k+1)}m\right)
  \end{split}
 \end{equation}
 and
 \begin{equation}
  \label{Eq:At}
  \begin{split}
   \At_\delta^m &= \sum_{j=0}^m (-1)^j \tbinom mj
                      \left(\tbinom{\delta/2+m-j-1      -(m-j)(d+1)-j(k+1)}m\right.  \\
                &     \qquad
                      \vphantom{%
                      \left(\tbinom{\delta/2+m-j-1      -(m-j)(d+1)-j(k+1)}m\right.}
                         {}-\tbinom{\delta/2+m-j  -(d+1)-(m-j)(d+1)-j(k+1)}m         \\
                &     \qquad
                         {}+\tbinom{\delta/2+m-j-1-(k+1)-(m-j)(d+1)-j(k+1)}m
                      \vphantom{                                            \left.
                            \tbinom{\delta/2+m-j-1-(r+1)-(m-j)(d+1)-j(k+1)}m\right)} \\
                &     \qquad
                                                                            \left.
                         {}-\tbinom{\delta/2+m-j-1-(r+1)-(m-j)(d+1)-j(k+1)}m\right).
  \end{split}
 \end{equation}

 Also, let
 \begin{equation}
  \label{Eq:B}
  \begin{split}
   B_{n,\delta}^m &= \sum_{j=0}^m (-1)^j \tbinom mj
                      \left(\tbinom{n-\delta/2+m-j        -(m-j)(d+1)-j(k+1)}m
                       \vphantom{%
                           -\tbinom{n-\delta/2+m-j-1-(r+1)-(m-j)(d+1)-j(k+1)}m}\right. \\
                  &   \qquad
                                                                               \left.
                       \vphantom{%
                            \tbinom{n-\delta/2+m-j        -(m-j)(d+1)-j(k+1)}m}
                         {}-\tbinom{n-\delta/2+m-j-1-(r+1)-(m-j)(d+1)-j(k+1)}m\right)
  \end{split}
 \end{equation}
 and
 \begin{equation}
  \label{Eq:Bt}
  \begin{split}
   \Bt_{n,\delta}^m &= \sum_{j=0}^m (-1)^j \tbinom mj
                      \left(\tbinom{n-\delta/2+m-j  -(m-j)(d+1)-j(k+1)}m
                       \vphantom{%
                           -\tbinom{n-\delta/2+m-j-1-(m-j)(d+1)-j(k+1)}m}\right. \\
                    & \qquad
                                                                         \left.
                       \vphantom{%
                            \tbinom{n-\delta/2+m-j  -(m-j)(d+1)-j(k+1)}m}
                         {}-\tbinom{n-\delta/2+m-j-1-(m-j)(d+1)-j(k+1)}m\right).
  \end{split}
 \end{equation}
 Using this notation, we may rewrite $\CCS_n^{-n}$ and $\CCS_n^{-n+2}$
 from our examples as
 \begin{align}
  \CCS_n^{-n}        & =A_0^0B_{n,0}^0, \notag \\
  \CCS_n^{-n+2}      & =A_2^1B_{n,2}^1+\At_2^0\Bt_{n,2}^1. \notag \\
  \intertext{Likewise,}
  \CCS_n^{-n+4}      & =A_4^1B_{n,4}^1+A_4^2B_{n,4}^2
                       +\At_4^0\Bt_{n,4}^1+\At_4^1\Bt_{n,4}^2. \notag \\
  \intertext{Continuing in the same way, we see that}
  \label{Eq:CCSdeltaL}
  \CCS_n^{-n+\delta} & =\sum_{m=0}^{\delta/2}
                        \left( A_{\delta}^m       B_{n,\delta}^m
                            +\At_{\delta}^{m-1} \Bt_{n,\delta}^m \right).
 \end{align}
 Observe that the upper limit of summation might be written as a quotient of
 $ \delta $ and $ {2(d+1)} $. In such case
 \begin{equation}
  \label{Eq:CCSdeltaL1}
  \CCS_n^{-n+\delta}=\sum_{m=0}^{\left\lceil \frac\delta{2(d+1)} \right\rceil}
                      \left( A_{\delta}^m       B_{n,\delta}^m
                          +\At_{\delta}^{m-1} \Bt_{n,\delta}^m \right).
 \end{equation}

 Note that summing $\CCS_n^{-n+\delta}$ in right slanting rows
 does not affect $B_{n,\delta}^m$ and $\Bt_{n,\delta}^m$.
 This affects only $A_\delta^m$ and $\At_\delta^m$.
 Indeed, from~\eqref{Eq:Crsrs_dkr1} it follows that index of summation,
 which subtracts from $n$ and from $\delta/2$,
 annihilates in~\eqref{Eq:B} and~\eqref{Eq:Bt}
 and remains in~\eqref{Eq:A} and~\eqref{Eq:At}.

 Now, consider $\sum_{i=d+1}^{k+1} A_{\delta-i}^m$. We have
 \begin{equation}
  \label{Eq:ArecurChain}
  \begin{split}
   & \begin{split}
      \sum_{i=d+1}^{k+1} A_{\delta-i}^m
       &=\sum_{j=0}^m       (-1)^j \tbinom mj \\
       &  \quad
       {}\times%
          \left(\tbinom{\delta/2+m-j  -(m-j)(d+1)-j(k+1)-(d+1)}m
           \vphantom{%
               -\tbinom{\delta/2+m-j-1-(m-j)(d+1)-j(k+1)-(d+1)}m}\right. \\
       &  \qquad
             {}-\tbinom{\delta/2+m-j-1-(m-j)(d+1)-j(k+1)-(d+1)}m         \\
       &  \qquad
      {}+       \tbinom{\delta/2+m-j  -(m-j)(d+1)-j(k+1)-(d+2)}m         \\
       &  \qquad
             {}-\tbinom{\delta/2+m-j-1-(m-j)(d+1)-j(k+1)-(d+2)}m         \\
       &  \qquad
         \cdots%
        +       \tbinom{\delta/2+m-j  -(m-j)(d+1)-j(k+1)-(k+1)}m         \\
       &  \qquad
                                                                 \left.
           \vphantom{%
                \tbinom{\delta/2+m-j  -(m-j)(d+1)-j(k+1)-(k+1)}m}
             {}-\tbinom{\delta/2+m-j-1-(m-j)(d+1)-j(k+1)-(k+1)}m\right)
     \end{split} \\
   & \begin{split}
        =\sum_{j=0}^m &     (-1)^j \tbinom mj
          \left(\tbinom{\delta/2+m-j  -(m-j)(d+1)-j(k+1)+1-(d+1)}{m+1}
           \vphantom{%
               -\tbinom{\delta/2+m-j-1-(m-j)(d+1)-j(k+1)  -(k+1)}{m+1}}\right.\\
                      & \qquad \quad
             {}-\tbinom{\delta/2+m-j  -(m-j)(d+1)-j(k+1)  -(k+1)}{m+1}        \\
                      & \qquad \quad
       {}-\left(\tbinom{\delta/2+m-j-1-(m-j)(d+1)-j(k+1)+1-(d+1)}{m+1}\right. \\
                      & \qquad \quad
                                                                      \left.
           \vphantom{%
                \tbinom{\delta/2+m-j  -(m-j)(d+1)-j(k+1)+1-(d+1)}{m+1}}
       {}-\left.\tbinom{\delta/2+m-j-1-(m-j)(d+1)-j(k+1)  -(k+1)}{m+1}\right)
                                                                      \right)
     \end{split} \\
   & \begin{split}
        &= \sum_{j=0}^m     (-1)^j \tbinom mj
          \left(\tbinom{\delta/2+m+1-j  -(m+1-j)(d+1)-j(k+1)}{m+1}\right.\\
        & \qquad \qquad
       {}-\left.\tbinom{\delta/2+m+1-j-1-(m+1-j)(d+1)-j(k+1)}{m+1}\right)\\
        & \quad
      {}+\sum_{j=1}^{m+1}   (-1)^j \tbinom m{j-1}
          \left(\tbinom{\delta/2+m+1-j  -(m+1-j)(d+1)-j(k+1)}{m+1}\right.\\
        & \qquad \qquad
       {}-\left.\tbinom{\delta/2+m+1-j-1-(m+1-j)(d+1)-j(k+1)}{m+1}\right)
     \end{split} \\
   & \begin{split}
        =\sum_{j=0}^{m+1} & (-1)^j \tbinom{m+1}j
          \left(\tbinom{\delta/2+m+1-j  -(m+1-j)(d+1)-j(k+1)}{m+1}\right.\\
                          & \qquad
       {}-\left.\tbinom{\delta/2+m+1-j-1-(m+1-j)(d+1)-j(k+1)}{m+1}\right).
     \end{split}
  \end{split}
 \end{equation}
 This means that
 \begin{equation*}
  \sum_{i=d+1}^{k+1} A_{\delta-i}^m=A_{\delta}^{m+1}.
 \end{equation*}
 Similarly,
 \begin{equation*}
  \sum_{i=d+1}^{k+1} \At_{\delta-i}^m=\At_{\delta}^{m+1}.
 \end{equation*}
 Using this result and using~\eqref{Eq:Crsrs_dkr1} again we see that
 \begin{equation*}
  \CCS_n^{n-\delta}=\sum_{m=0}^{\delta/2}
                     \left( A_{\delta}^{m+1} B_{n,\delta}^m
                         +\At_{\delta}^m   \Bt_{n,\delta}^m \right),
                    \quad \delta > 0
 \end{equation*}
 and like~\eqref{Eq:CCSdeltaL1}, we may rewrite
 \begin{equation*}
  \CCS_n^{n-\delta}=\sum_{m=0}^{\left\lceil \frac\delta{2(d+1)} \right\rceil}
                     \left( A_{\delta}^{m+1} B_{n,\delta}^m
                         +\At_{\delta}^m   \Bt_{n,\delta}^m \right),
                    \quad \delta > 0.
 \end{equation*}

 For $\chrg=-n+\delta$,
 we have from recursion relation~\eqref{Eq:Cc} that
 \begin{equation}
  \label{Eq:Clsrs_dkr1}
  \CCS_n^{-n+\delta}=\CCS_{n-(d+1)}^{n-(d+1)-\delta}%
                    +\CCS_{n-(d+2)}^{n-(d+2)-\delta}%
                    +\dots%
                    +\CCS_{n-(k+1)}^{n-(k+1)-\delta}.
 \end{equation}

 Similarly, we can see that summing $\CCS_n^{n-\delta}$ in left slanting rows
 does not affect $A_\delta^m$ and $\At_\delta^m$.
 This affects $B_{n,\delta}^m$ and $\Bt_{n,\delta}^m$.
 Namely, from~\eqref{Eq:Clsrs_dkr1} it follows that index of summation
 subtracts only from $n$.

 Consider $\sum_{i=d+1}^{k+1} B_{n-i,\delta}^m$.
 Applying a reasoning chain like~\eqref{Eq:ArecurChain} yields
 \begin{equation*}
  \begin{split}
   \sum_{i=d+1}^{k+1} & B_{n-i,\delta}^m
     =\sum_{j=0}^{m+1} (-1)^j \tbinom{m+1}j \\
                      & \qquad
    {}\times%
       \left(\tbinom{n-\delta/2+m+1-j        -(m+1-j)(d+1)-j(k+1)}{m+1}\right.\\
                      &
    {}-\left.\tbinom{n-\delta/2+m+1-j-1-(r+1)-(m+1-j)(d+1)-j(k+1)}{m+1}\right).
  \end{split}
 \end{equation*}
 This means that
 \begin{equation*}
  \sum_{i=d+1}^{k+1} B_{n-i,\delta}^m=B_{n,\delta}^{m+1}.
 \end{equation*}
 Similarly,
 \begin{equation*}
  \sum_{i=d+1}^{k+1} \Bt_{n-i,\delta}^m=\Bt_{n,\delta}^{m+1}.
 \end{equation*}
 Using this result and using~\eqref{Eq:Clsrs_dkr1} we obtain
 \begin{equation}
  \label{Eq:CCSdeltaLL}
  \CCS_n^{-n+\delta}=\sum_{m=0}^{\delta/2}
                      \left( A_{\delta}^{m+1} B_{n,\delta}^{m+1}
                          +\At_{\delta}^m   \Bt_{n,\delta}^{m+1} \right),
                     \quad \delta > 0.
 \end{equation}
 This proves the inductive hypothesis~\eqref{Eq:CCSdeltaL}.

 Like~\eqref{Eq:CCSdeltaL1}, we may rewrite~\eqref{Eq:CCSdeltaLL} as
 \begin{equation}
  \label{Eq:CCSdeltaLL1}
  \CCS_n^{-n+\delta}=\sum_{m=0}^{\left\lceil \frac\delta{2(d+1)} \right\rceil}
                      \left( A_{\delta}^{m+1} B_{n,\delta}^{m+1}
                          +\At_{\delta}^m   \Bt_{n,\delta}^{m+1} \right),
                     \quad \delta > 0.
 \end{equation}

 Substituting $n+\chrg$ for $\delta$ in~\eqref{Eq:CCSdeltaL1},
 we get~\eqref{Eq:CCSsigmaL} at the bottom of the page.

 \begin{figure*}[!b]
  \normalsize
  \vspace*{4pt}
  \hrulefill
  \begin{multline}
  \label{Eq:CCSsigmaL}
   \CCS_n^\chrg=\sum_{m=0}^{\left\lceil \tfrac{n+\chrg}{2(d+1)}
                            \right\rceil}
          \left(\left(\sum_{j=0}^m (-1)^j \tbinom mj
           \left(   \tbinom{\tfrac{n+\chrg}2        -md-j\rl2}m
                   -\tbinom{\tfrac{n+\chrg}2-1      -md-j\rl2}m\right)
                \right)
          \right.\\
        {}\times\left(\sum_{j=0}^m (-1)^j \tbinom mj
           \left(   \tbinom{\tfrac{n-\chrg}2        -md-j\rl2}m
                   -\tbinom{\tfrac{n-\chrg}2-1-(r+1)-md-j\rl2}m\right)
                \right)\\
             {}+\left(\sum_{j=0}^{m-1}(-1)^j \tbinom{m-1}j
           \left(   \tbinom{\tfrac{n+\chrg}2-1      -(m-1)d-j\rl2}{m-1}
                   -\tbinom{\tfrac{n+\chrg}2  -(d+1)-(m-1)d-j\rl2}{m-1}\right.
                \right.\\
                \left.\vphantom{%
                      \sum_{j=0}^{m-1}(-1)^j \tbinom{m-1}j
           \left(   \tbinom{\tfrac{n+\chrg}2-1      -(m-1)d-j\rl2}{m-1}
                   -\tbinom{\tfrac{n+\chrg}2  -(d+1)-(m-1)d-j\rl2}{m-1}\right.}
           \left.{}+\tbinom{\tfrac{n+\chrg}2-1-(k+1)-(m+1)d-j\rl2}{m-1}
                   -\tbinom{\tfrac{n+\chrg}2-1-(r+1)-(m-1)d-j\rl2}{m-1}\right)
                \right)\\
         \left.
        {}\times\left(\sum_{j=0}^m (-1)^j \tbinom mj
           \left(   \tbinom{\tfrac{n-\chrg}2        -md-j\rl2}m
                   -\tbinom{\tfrac{n-\chrg}2-1      -md-j\rl2}m\right)
                \right)
         \right)
  \end{multline}
 \end{figure*}

 \section{Generating Functions for Enumerating 
  the Constant-Weight and Constant-Charge Sequences}
  \label{Sec:GenFunct}

 From triangle table (see Table~\ref{Tab:ChrgDistr}(\alph{MYtempcnt}))
 it follows that there exist two types of sequences
 $ \CWS_n^\nu $ and $ \CCS_n^\chrg $.
 The infinite sequences of the first type are
 $ \CWS_0^\nu , \CWS_1^\nu , \dots $ then
 $ \CCS_0^\chrg , \CCS_2^\chrg , \dots $ or $ \CCS_1^\chrg , \CCS_3^\chrg , \dots $.
 The finite sequences of the second type are
 $ \CWS_n^1 , \CWS_n^2 , \dots , \CWS_n^n $ and
 $ \CCS_n^{ -n } , \CCS_n^{ -n + 2 } , \dots , \CCS_n^n $.
 Kolesnik and Krachkovsky described in~\cite{Kolesnik91}
 a recursive calculation of generating function for enumerating
 constant-weight sequences of the first type.
 Lee in~\cite{Lee88} suggested a direct method for the same.
 Here we obtain the generating functions of the both types.

 \subsection{Generating Function of Sequence %
  $ \CWS_0^\nu , \CWS_1^\nu , \dots $}

 Consider infinite sequence
 $\CWS_0^\nu, \CWS_1^\nu, \CWS_2^\nu, \dots$.
 Define generating function of this sequence as
 \[
  \CWS^\nu(t)=\sum_{n=0}^\infty \CWS_n^\nu t^n, \quad t \in \mathbb R.
 \]
 Substituting~\eqref{Eq:Adkr} for $\CWS_n^\nu$ here
 and changing the order of summation we have that
 \begin{equation*}
  \begin{split}
   \CWS^\nu(t) &= \sum_{j=0}^{\nu-1} (-1)^j \tbinom{\nu-1}j \\
               &  \quad
                 {}\times%
                   \left(   \sum_{\substack{n=(\nu-1)+1+(\nu-1)d \\
                                            {}+j\rl2}}^\infty%
                    \tbinom{n-1-(\nu-1)d-j\rl2}{\nu-1} t^n
                   \right. \\
               &  \qquad
                   \left.{}-\sum_{\substack{n=(\nu-1)+1+(\nu-1)d \\
                                            {}+(r+1)+j\rl2}}^\infty %
                    \tbinom{n-1-(\nu-1)d-(r+1)-j\rl2}{\nu-1} t^n
                   \right).
  \end{split}
 \end{equation*}
 By using the fact that
 $\sum\limits_{n=\nu+a}^\infty \binom{n-a}\nu t^n%
  =\left(t^\nu / (1-t)^{\nu+1}\right)t^a$ we get
 \begin{equation}
  \label{Eq:GFAn}
  \CWS^\nu(t)=\frac{t^{\nu-1}}{(1-t)^\nu}
              t^{1+(\nu-1)d}(1-t^{r+1})(1-t^{\rl1})^{\nu-1}.
 \end{equation}

 \subsection{Generating Function of Sequence %
  $ \CWS_n^1 , \CWS_n^2 , \dots , \CWS_n^n $}

 Consider a finite power series
 \[
  \CWS_n(y)=\sum_{\nu=1}^n \CWS_n^\nu y^\nu, \quad y \in \mathbb R.
 \]
 We can obtain $\CWS_n(y)$ in closed form by the following way.
 Consider a formal power series
 \[
  \CWS(t,y)=\sum_{\nu=1}^\infty \CWS^\nu(t) y^\nu.
 \]
 We can achieve convergence of this series by assuming $t$ arbitrarily small.
 Rewrite~\eqref{Eq:GFAn} as
 \[
  \CWS^\nu(t)=\left(\frac{t^{d+1}(1-t^{\rl1})}{1-t}\right)^{\nu-1}
              \frac{t(1-t^{r+1})}{1-t}.
 \]
 Then we have
 \begin{align}
  \CWS(t,y) &=\frac{yt(1-t^{r+1})}{1-t}
               \sum_{\nu=1}^\infty \left(\frac{yt^{d+1}(1-t^{\rl1})}
                                              {1-t}
                                   \right)^{\nu-1} \notag \\
  \label{Eq:GFA}
            &=\frac{yt(1-t^{r+1})}
                   {1-t-yt^{d+1}+yt^{k+2}}.
 \end{align}
 By using the fact that
 $d \CWS(t,y)/dt
  =\sum\limits_{n=1}^\infty n \CWS_n(y) t^{n-1}$ we obtain
 \[
  \CWS_n(y)=\frac 1{n!} \lim_{t \to 0} \frac{ d^n \CWS(t,y)}
                                            {dt^n}.
 \]
 From Cauchy's integral representation, we write
 \[
  \frac{ d^n \CWS(t,y)}
       {dt^n}          =\frac{n!}{2\pi i}\oint_\Gamma
                         \frac{y\tau(1-\tau^{r+1})}
                              {\left(1-\tau-y\tau^{d+1}+y\tau^{k+2}\right)
                                (\tau-t)^{n+1}} d\tau.
 \]
 Then
 \begin{equation}
  \label{Eq:GFAnuI}
  \CWS_n(y)=\frac{y}{2\pi i}\oint_\Gamma
             \frac{1-\tau^{r+1}}
                  {\left(1-\tau-y\tau^{d+1}+y\tau^{k+2}\right)\tau^n} d\tau,
 \end{equation}
 where $\Gamma$ is a counterclockwise simple closed contour
 surrounding the origin of the complex plane,
 small enough to avoid any other poles
 of integrand in~\eqref{Eq:GFAnuI}.
 The application of the residue theorem yields
 \begin{equation}
  \label{Eq:GFAnu}
  \CWS_n(y)=\sum_{j=1}^{k+2} \frac{\tau_j^{r+1}-1}
                                  {\tau_j^n\displaystyle
                                   \prod_{\substack{m=1,\\ m \neq j}}^{k+2}
                                   (\tau_j-\tau_m)},
 \end{equation}
 where $\tau_1$, $\tau_2$, $\dots$, $\tau_{k+2}$ are roots of
 $1-\tau-y\tau^{d+1}+y\tau^{k+2}$ and $\tau_1=1$.

 \subsection{Generating Function of Sequences %
     $ \CCS_0^\chrg , \CCS_2^\chrg , \dots $ %
  or $ \CCS_1^\chrg , \CCS_3^\chrg , \dots $}

 Similarly, consider an infinite sequence
 $\CCS_0^\chrg, \CCS_1^\chrg, \CCS_2^\chrg, \dots$.
 In such case, we define generating function as
 \begin{equation}
  \label{Eq:Ct1}
  \CCS^\chrg(t)=\sum_{\substack{n=0 \\ n\ \mathrm{even}}}^\infty
                                      \CCS_n^\chrg t^{n/2},
  \quad \text{or} \quad
  \CCS^\chrg(t)=\sum_{\substack{n=1 \\ n\ \mathrm{odd}}}^\infty
                                      \CCS_n^\chrg t^{(n-1)/2},
 \end{equation}
 where $t \in \mathbb R$.

 It is easily shown that~\eqref{Eq:CCSsigmaL} can be written
 as a sum of $12$ terms with proper signs.
 \begin{equation}
  \label{Eq:CCSi}
  \CCS_n^\chrg=\sum_{\iota=1}^{12}(-1)^{\lfloor \iota/2 \rfloor}
                \CCSi_n^\chrg.
 \end{equation}
 Define these terms $\CCSi_n^\chrg$ as
 \begin{equation}
  \label{Eq:CdkrI}
  \begin{split}
   \CCSi_n^\chrg &= \sum_{m=0}^{\left\lceil
                                      \tfrac{n+\chrg}{2(d+1)}
                                \right\rceil}
               \biggl(\sum_{j=0}^{m-\fst} (-1)^j \tbinom{m-\fst}j \\
                 & \qquad \qquad \qquad
          {}\times
          \tbinom{\tfrac{n+\chrg}2-\ats-md-j\rl2-\fst}{m-\fst}
               \biggr) \\
                 & \qquad \quad
       {}\times \biggl(\sum_{j=0}^m (-1)^j \tbinom mj
          \tbinom{\tfrac{n-\chrg}2-\bts-md-j\rl2     } m
               \biggr),
  \end{split}
 \end{equation}
 where
 \[
  \ats=\begin{cases}
        0,       & I=1,  2,  7,  8,\\
        1,       & I=3,  4,\\
        -d,      & I=5,  6,\\
        (k+1)-d, & I=9,  10,\\
        (r+1)-d, & I=11, 12,
       \end{cases}
 \]
 \[
  \bts=\begin{cases}
        0,       & I=1,  3,  5,  7,  9, 11,\\
        1+(r+1), & I=2,  4,\\
        1,       & I=6,  8, 10, 12,
       \end{cases}
 \]
 and
 \[
  \fst=\begin{cases}
        0, & I=1,  2,  3,  4,\\
        1, & I=5,  6,  7,  8,  9, 10, 11, 12.
       \end{cases}
 \]
 Without loss of generality we will consider a term of~\eqref{Eq:CCSi}
 as $ \CCS_n^\chrg $
 and assume that $ \CCS_n^\chrg := \CCSi_n^\chrg $ below in this section.
 Of course, this implies using~\eqref{Eq:CCSi} when calculating.
 \renewcommand{\CCSi}{\CCS}
 \renewcommand{\rl}[1]{q}              

 We can expand the inner product of two series using double sums as follows:
 \begin{equation*}
  \begin{split}
   &        \biggl(\sum_{j=0}^{m-\fst} (-1)^j \tbinom{m-\fst}j
       \tbinom{\tfrac{n+\chrg}2-\ats-md-j\rl2-\fst}{m-\fst}
            \biggr) \\
   & \qquad \qquad
    {}\times\biggl(\sum_{j=0}^m        (-1)^j \tbinom m      j
       \tbinom{\tfrac{n-\chrg}2-\bts-md-j\rl2     } m
            \biggr) \\
   &=
                   \sum_{u=0}^{m-\fst} \sum_{v=0}^m (-1)^{u+v}
                                                  \tbinom{m-\fst}u
                                                  \tbinom m      v \\
   & \qquad \qquad \qquad
    {}\times \tbinom{\tfrac{n+\chrg}2-\ats-md-u\rl2-\fst}{m-\fst}
             \tbinom{\tfrac{n-\chrg}2-\bts-md-v\rl2     } m;
  \end{split}
 \end{equation*}
 then interchanging the sums, we get
 \begin{equation}
  \label{Eq:Ct5}
  \CCSi^\chrg(t)=\sum_{m=0}^\infty
    \left(\sum_{u=0}^{m-\fst} \sum_{v=0}^m (-1)^{u+v}
     \binom{m-\fst}u
     \binom m      v
     g(t)
    \right),
 \end{equation}
 where by $g(t)$ we denote the inner sum.
 \begin{equation}
  \label{Eq:g1}
  g(t)=\sum_{\substack{n=\eo            \\
                       n\ \mathrm{even} \\
                          \text{or}     \\
                       n\ \mathrm{odd}}}^\infty
               \tbinom{\tfrac{n+\chrg}2-\ats-md-u\rl2-\fst}{m-\fst}%
               \tbinom{\tfrac{n-\chrg}2-\bts-md-v\rl2     } m
                                                            t^{(n-\eo)/2},
 \end{equation}
 where $\eo=\chrg \bmod 2$.

 Further, we will derive the generating function using arguing style
 of orthogonal polynomials theory~\cite{Szego75}.

 Let $G(n)=\tbinom{\tfrac{n+\chrg}2-\ats-md-u\rl2-\fst}{m-\fst}%
           \tbinom{\tfrac{n-\chrg}2-\bts-md-v\rl2     } m
           t^{(n-\eo)/2}$,                             $m-\fst \geq 0$.
 The sequence $G(n)$ may content leading zero elements.
 In this case we can not write a ratio between consecutive terms.
 To prevent this we can use the following rule
 \[
  \sum_{j=0}^\infty
   \tbinom{j-A}B%
   \tbinom{j-C}D
      =\begin{cases}
        \displaystyle \sum_{j=A+B}^\infty
         \tbinom{j-A}B
         \tbinom{j-C}D,
                      & A+B-C \geq D, \\
        \displaystyle \sum_{j=C+D}^\infty
         \tbinom{j-A}B
         \tbinom{j-C}D,
                      & C+D-A \geq B.
       \end{cases}
 \]
 Let $\alpha$ be the $A+B-C-D$ in the conditionals above. In our case
 $\alpha=-\chrg+\ats-\bts+(u-v)\rl2$ and
 we can rewrite~\eqref{Eq:g1} as
 \begin{equation}
  \label{Eq:gt1}
  \begin{split}
   g(t) &= t^{m(d+1)+a} \\
        & \qquad
          {}\times%
           \sum_{j=0}^\infty
            \binom{m-\beta+j}{m-\beta}%
            \binom{m-\fst+\beta+|\alpha|+j}{m-\fst+\beta} t^j,
  \end{split}
 \end{equation}
 where
 \[
  a=\begin{cases}
     \fmc+\ats+u\rl2, & \alpha \geq 0, \\
     \fpc+\bts+v\rl2, & \text{otherwise}
    \end{cases}
 \]
 and
 \[
  \beta=\begin{cases}
         \fst, & \alpha \geq 0, \\
         0,    & \text{otherwise}.
        \end{cases}
 \]
 Now, consider the sum in~\eqref{Eq:gt1}.
 By $G(j)$ denote a term of this series
       $\binom{m-\beta+j}{m-\beta}%
        \binom{m-\fst+\beta+|\alpha|+j}{m-\fst+\beta} t^j$
 and write the term ratio as
 \[
  \begin{split}
   \frac{G(j+1)}{G(j)} &=
     \frac{\tbinom{m-\beta+j+1}{m-\beta}%
           \tbinom{m-\fst+\beta+|\alpha|+j+1}{m-\fst+\beta} t^{j+1}}
          {\tbinom{m-\beta+j  }{m-\beta}%
           \tbinom{m-\fst+\beta+|\alpha|+j  }{m-\fst+\beta} t^{j}} \\
   &=\frac{(j+2)_{m-\beta} (|\alpha|+j+2)_{m-\fst+\beta} t}
          {(j+1)_{m-\beta} (|\alpha|+j+1)_{m-\fst+\beta}} \\
   &=\frac{(m-\beta+j+1)(m-\fst+\beta+|\alpha|+j+1) t}
          {        (j+1)(             |\alpha|+j+1)} ,
  \end{split}
 \]
 where $(a)_j$ denote the Pochhammer-symbol which is defined by
 $(a)_0=1$ and $(a)_j=a(a+1)(a+2)\cdots(a+j-1)$, $j=1,2,3,\dots$
 Obviously that $G(0)=\binom{m-\fst+\beta+|\alpha|}{m-\fst+\beta}$
 and hence,
 \newcommand{\GO}{G(0)}
 \[
  G(j)=\GO
           \frac{(m-\beta+1)_j (m-\fst+\beta+|\alpha|+1)_j}
                {                           (|\alpha|+1)_j}
           \frac{t^j}{j!}.
 \]
 By summing $G(j)$ we get the hypergeometric series as follows:
 \begin{equation*}
  \begin{split}
   g(t) &= t^{m(d+1)+a} \GO       \\
        & \qquad \qquad
           {}\times \hyp{m-\beta+1, m-\fst+\beta+|\alpha|+1}
                        {                        |\alpha|+1}{t}.
  \end{split}
 \end{equation*}
 If we apply the Pfaff-Euler transformation formula \cite{abramowitz+stegun},
 then we obtain the next terminating series
 \[
  g(t)=\frac{t^{m(d+1)+a}}{(1-t)^{2m-\fst+1}}
       \GO
       \hyp{-m+\fst-\beta, |\alpha|+\beta-m}{|\alpha|+1}{t}.
 \]
 By several transformations we get
 \begin{equation*}
  \begin{split}
   & g(t)=\frac{t^{m(d+1)+a}}{(1-t)^{2m-\fst+1}}
          \GO
                   \sum_{j=0}^{m-\fst+\beta}%
                 \binom{m-\fst+\beta}j \\
   &
          {}\times%
          \frac{(m-\beta-|\alpha|)(m-\beta-|\alpha|-1)\cdots%
                (m-\beta-|\alpha|-j+1)}
               {(|\alpha|+1)(|\alpha|+2)\cdots(|\alpha|+1+j-1)} t^j
  \end{split}
 \end{equation*}
 then, using the distributive law, we obtain
 \begin{equation*}
  \begin{split}
   & g(t)=\frac{t^{m(d+1)+a}}{(1-t)^{2m-\fst+1}}
          \frac 1{(m-\fst+\beta)!}
                   \sum_{j=0}^{m-\fst+\beta} \binom{m-\fst+\beta}j \\
   &
          {}\times%
          (m-\fst+\beta+|\alpha|)(m-\fst+\beta+|\alpha|-1)\cdots%
          (|\alpha|+j+1) \\
   &
          {}\times%
          (m-\beta-|\alpha|)(m-\beta-|\alpha|-1)\cdots(m-\beta-|\alpha|-j+1)
                                                               t^j.
  \end{split}
 \end{equation*}
 Now, our goal is to apply the Leibniz rule to this finite series.
 We have a power series in one variable $t$.
 But, we need a power series in two variables
 because we have rising and falling factorial powers here.
 Recall the M\"obius transformation and suppose that
 \begin{equation}
  \label{Eq:tx1}
  t=\frac{x-1}
         {x+1}, \quad \parbox[c]{.36\linewidth}{%
                       (assume $x \in \mathbb R$ till the end of this section)}
 \end{equation}
 \newcommand{\tx}{t}
 and rewrite
 \begin{equation*}
  \begin{split}
   g(t) &= \tx^{m(d+1)+a-|\alpha|}%
                 \frac{(x+1)^{m-\fst+\beta+1}}{2^{2m-\fst+1}} \\
        & \qquad
           {}\times%
           \frac 1{(m-\fst+\beta)!}
                      \sum_{j=0}^{m-\fst+\beta} \binom{m-\fst+\beta}j \\
        & \quad
           {}\times%
           (m-\fst+\beta+|\alpha|)(m-\fst+\beta+|\alpha|-1)\cdots \\
        & \qquad
           {}\times%
           (m-\fst+\beta+|\alpha|-(m-\fst+\beta-j)+1) \\
        & \qquad \qquad
           {}\times%
           (x-1)^{m-\fst+\beta+|\alpha|-(m-\fst+\beta-j)} \\
        & \qquad
           {}\times%
           (m-\beta-|\alpha|)(m-\beta-|\alpha|-1)\cdots \\
        & \qquad \qquad
           {}\times%
           (m-\beta-|\alpha|-j+1)
           (x+1)^{m-\beta-|\alpha|-j}.
  \end{split}
 \end{equation*}
 So, we obtain $(m-\fst+\beta-j)$th derivative of $(x+1)^{m-\fst+\beta+1}$
 and $(j)$th derivative of                        $(x+1)^{m     -\beta-|\alpha|}$.
 \begin{equation*}
  \begin{split}
   & g(t)=\tx^{m(d+1)+\bar a}%
                \frac{(x+1)^{m-\fst+\beta+1}}{2^{2m-\fst+1}}%
          \frac 1{(m-\fst+\beta)!} \\
   & \quad
          {}\times \sum_{j=0}^{m-\fst+\beta} \binom{m-\fst+\beta}j
          \frac{ d^{m-\fst+\beta-j}(x-1)^{m-\fst+\beta+|\alpha|}}
               {dx^{m-\fst+\beta-j}} \\
   & \qquad \qquad \qquad \qquad \qquad \qquad \qquad \quad
          {}\times%
          \frac{d^j                (x+1)^{m     -\beta-|\alpha|}}
               {dx^j},
  \end{split}
 \end{equation*}
 where $\bar a=a-|\alpha|$.
 Finally, we have the Rodrigues' formula
 \begin{equation*}
  \begin{split}
   g(t) &= \tx^{\bar a}%
                 \frac{(x+1)^{\beta+1-\fst}}{2^{1-\fst}} \bd^m%
           \frac 1{(m-\fst+\beta)!} \\
        & \qquad
           {}\times
           \frac{ d^{m-\fst+\beta}
                    \left(\left(\dfrac{x-1}{x+1}\right)^{|\alpha|+\beta}
                                \dfrac{(x^2-1)^m}{(x-1)^\fst}\right)}
                {dx^{m-\fst+\beta}},
  \end{split}
 \end{equation*}
 where
 \begin{equation}
  \label{Eq:b}
  \bd=\left(\frac{x-1}{x+1}\right)^{d+1}\frac{x+1}4
     =\frac{t^{d+1}}
           {2(1-t)}.
 \end{equation}
 Since we suppose~\eqref{Eq:tx1},
 then by
 $\tau = \frac{\xi-1}
              {\xi+1}$ we denote the similar mapping in the complex plane.
 \newcommand{\tauxi}{\tau}
 From Cauchy's integral representation, we write
 \begin{equation}
  \label{Eq:gxInt}
  \begin{split}
   g(t) &= \tx^{\bar a}%
                 \frac{(x+1)^{\beta+1-\fst}}{2^{1-\fst}} \bd^m \\
        & \qquad \qquad
           {}\times
           \frac 1{2\pi i}\oint_\gamma
           \frac{\tauxi^{|\alpha|+\beta}%
                                    (\xi^2-1)^m}
                {(\xi-1)^\fst(\xi-x)^{m-\fst+\beta+1}} d\xi,
  \end{split}
 \end{equation}
 where $\gamma$ is some Jordan curve about the point
 $x = \frac{1+t}
           {1-t}$; this point lies on the real axis.

 Here, we conclude the principal part of our derivation.
 This sequence of transformation also is need for the sequel, at least twice.

 Now, recall~\eqref{Eq:Ct5}.
 Interchanging the order of summation and integration;
 then interchanging the order of summation (with proper justification) yields
 \begin{equation*}
  \begin{split}
   \CCSi^\chrg(t) &= \left(\frac{x+1}2\right)^{1-\fst}
                    \frac 1{2\pi i}\oint_\gamma
        \left(\sum_{u=0}^\infty \sum_{v=0}^\infty
         (-1)^{u+v}
        \vphantom{%
               \tx^{\bar a}
               (x+1)^\beta
         \frac{\tauxi^{|\alpha|+\beta}}
              {(\xi-1)^\fst(\xi-x)^{\beta-\fst+1}}
                \sum_{m=\max(u+\fst,v)}^\infty
         \binom{m-\fst}u
         \binom m      v
         s^m}
        \right. \\
                  & \qquad
       {}\times%
               \tx^{\bar a}
               (x+1)^\beta
         \frac{\tauxi^{|\alpha|+\beta}}
              {(\xi-1)^\fst(\xi-x)^{\beta-\fst+1}} \\
                  & \qquad \qquad
        \left.\vphantom{%
              \sum_{u=0}^\infty \sum_{v=0}^\infty
         (-1)^{u+v}
               \tx^{\bar a}
               (x+1)^\beta
         \frac{\tauxi^{|\alpha|+\beta}}
              {(\xi-1)^\fst(\xi-x)^{\beta-\fst+1}}}
       {}\times%
                \sum_{m=\max(u+\fst,v)}^\infty
         \binom{m-\fst}u
         \binom m      v
         s^m
        \right) d\xi,
  \end{split}
 \end{equation*}
 where
 \begin{equation}
  \label{Eq:s}
  s=\bd \frac{\xi^2-1}
             {\xi-x}.
 \end{equation}
 The next step is interchanging the outer sums order for diagonal summing as
 $\sum_{u=0}^{m-1} \sum_{v=0}^m X(u,v)=%
  \sum_{u=0}^{m-1} \sum_{w=0}^{m-1-u} X(u+w,w)+%
  \sum_{v=1}^m     \sum_{w=0}^{m  -v} X(w,v+w)$.
 Now, we have to redefine $u:=u+w$ and $v:=w$ for the first  double sum;
 also                     $u:=w$ and $v:=v+w$ for the second double sum.
 Then, by $\alpha_u$, $\alpha_v$, $a_u$, $a_v$, $\beta_u$, and $\beta_v$
 we denote $\alpha$, $\bar a - w\rl2$, and $\beta$
 for the first and for the second double sums as shown in Table~\ref{Tab:DiagSum}.
 \begin{table}[!t]
  \renewcommand{\arraystretch}{1.3}
  \caption{Nomenclature for~\eqref{Eq:Cx8}}
  \label{Tab:DiagSum}
  \centering
  \tabcolsep=0.25em
  \begin{tabular}{c|c}
   The first double sum:             & The second double sum:              \\
   $\sum_{u=0}^\infty
    \sum_{w=0}^\infty$               & $\sum_{v=1}^\infty
                                        \sum_{w=0}^\infty$
                          $\vphantom{ \displaystyle \sum_S }$                \\
   \hline
   \parbox[b]{.48\linewidth}{%
    \begin{equation}
     \label{Eq:alphau}
     \alpha_u=-\chrg+\ats-\bts+u\rl2,
    \end{equation}}                  & \parbox[b]{.48\linewidth}{%
                                        \begin{equation}
                                         \label{Eq:alphav}
                                         \alpha_v=-\chrg+\ats-\bts-v\rl2,
                                        \end{equation}}                    \\
   \parbox[b]{.48\linewidth}{%
    \begin{equation}
     \label{Eq:au}
     a_u=\fpc+\bts+\begin{cases}
                    0,        & \alpha_u \!\geq 0, \\
                    \alpha_u, & \text{else},
                   \end{cases}
    \end{equation}}                  & \parbox[b]{.48\linewidth}{%
                                        \begin{equation}
                                         \label{Eq:av}
                                         \;
                                         a_v=\fmc+\ats-\begin{cases}
                                                        \alpha_v, & \!\!
                                                                    \alpha_v \!\geq 0, \\
                                                        0,        & \!\!
                                                                    \text{else},
                                                       \end{cases}
                                        \end{equation}}                    \\
   such that $\bar a = w\rl2+a_u$.   & such that $\bar a = w\rl2+a_v$.     \\
   \parbox[b]{.48\linewidth}{%
    \begin{equation}
     \label{Eq:betau}
     \beta_u=\begin{cases}
              \fst, & \alpha_u \geq 0, \\
              0,    & \text{else},
             \end{cases}
    \end{equation}}                  & \parbox[b]{.48\linewidth}{%
                                        \begin{equation}
                                         \label{Eq:betav}
                                         \beta_v=\begin{cases}
                                                  \fst, & \alpha_v \geq 0 \\
                                                  0,    & \text{else}.
                                                 \end{cases}
                                        \end{equation}}
  \end{tabular}
 \end{table}
 Similarly, denote by $f_{u,w}(s)$ and by $f_{v,w}(s)$ the inner sums over $m$
 and rewrite the result as
 \begin{equation}
  \label{Eq:Cx8}
  \begin{split}
   & \CCSi^\chrg(t)=\left(\frac{x+1}2\right)^{1-\fst} \\
   & \qquad \qquad \qquad
      {}\times%
                   \frac 1{2\pi i}\oint_\gamma
       \left(\sum_{u=0}^\infty (-1)^u \sum_{w=0}^\infty
              \tx^{a_u+w\rl2}
             \vphantom{%
              (x+1)^{\beta_u}
      {}\times%
        \frac{\tauxi^{|\alpha_u|+\beta_u}}
             {(\xi-1)^\fst(\xi-x)^{\beta_u-\fst+1}}
        f_{u,w}(s)}
       \right. \\
   & \qquad \qquad
       \left.\vphantom{%
             \sum_{u=0}^\infty (-1)^u \sum_{w=0}^\infty
              \tx^{a_u+w\rl2}}
      {}\times%
              (x+1)^{\beta_u}
        \frac{\tauxi^{|\alpha_u|+\beta_u}}
             {(\xi-1)^\fst(\xi-x)^{\beta_u-\fst+1}}
        f_{u,w}(s)
       \right. \\
   & \qquad \qquad \qquad \qquad
       \left.
           {}+\sum_{v=1}^\infty (-1)^v \sum_{w=0}^\infty
              \tx^{a_v+w\rl2}
             \vphantom{%
              (x+1)^{\beta_v}
      {}\times%
        \frac{\tauxi^{|\alpha_v|+\beta_v}}
             {(\xi-1)^\fst(\xi-x)^{\beta_v-\fst+1}}
        f_{v,w}(s)}
       \right. \\
   & \quad
       \left.\vphantom{%
           {}+\sum_{v=1}^\infty (-1)^v \sum_{w=0}^\infty
              \tx^{a_v+w\rl2}}
      {}\times%
              (x+1)^{\beta_v}
        \frac{\tauxi^{|\alpha_v|+\beta_v}}
             {(\xi-1)^\fst(\xi-x)^{\beta_v-\fst+1}}
        f_{v,w}(s)
       \right) d\xi,
  \end{split}
 \end{equation}
 where
 \begin{equation}
  \label{Eq:fuDef}
  f_{u,w}(s)=\sum_{m=u+w+\fst}^\infty
     \binom{m-\fst}{u+w}
     \binom m         w
     s^m, \quad u \geq 0,
 \end{equation}
 and
 \begin{equation}
  \label{Eq:fvDef}
  f_{v,w}(s)=\sum_{m=v+w}^\infty
     \binom{m-\fst}w
     \binom m   {v+w}
     s^m, \quad v \geq 1.
 \end{equation}
 It is easily shown that~\eqref{Eq:fuDef} and~\eqref{Eq:fvDef}
 are, in principal, similar to~\eqref{Eq:gt1}.
 Arguing as above (see~\eqref{Eq:gt1}~\dots~\eqref{Eq:gxInt}) we obtain
 \begin{equation}
  \label{Eq:fuwzi2}
  \begin{split}
   f_{u,w}(z) &= \left(\frac{z-1}{z+1}\right)^w%
                       \frac{(z+1)^{w+1}}{2^{u+2w+1}} \\
              &
                 {}\times
                 \frac 1{2\pi i}\oint_\Gamma
                 \frac{(\zeta-1)^u
                         \left(\dfrac{\zeta-1}{\zeta+1}\right)^\fst
                              (\zeta^2-1)^w}
                      {(\zeta-z)^{w+1}} d\zeta
  \end{split}
 \end{equation}
 and
 \begin{equation}
  \label{Eq:fvwzi2}
  \begin{split}
   f_{v,w}(z) &= \left(\frac{z-1}{z+1}\right)^{w+\fst}%
                       \frac{(z+1)^{w+1}}{2^{v+2w+1}} \\
              &
                 {}\times
                 \frac 1{2\pi i}\oint_\Gamma
                 \frac{(\zeta-1)^v
                         \left(\dfrac{\zeta+1}%
                                     {\zeta-1}\right)^\fst
                              (\zeta^2-1)^w}
                      {(\zeta-z)^{w+1}} d\zeta,
  \end{split}
 \end{equation}
 where $\Gamma$ is a closed contour surrounding the point $\zeta=z$.
 Here we suppose that
 \begin{equation*}
  s=\frac{z-1}
         {z+1}, \quad \parbox[c]{.36\linewidth}{%
                       (assume $z \in \mathbb C$ till the end of this section).}
 \end{equation*}

 Substituting~\eqref{Eq:fuwzi2}, and~\eqref{Eq:fvwzi2} for
 $f_{u,w}(s)$ and
 $f_{v,w}(s)$ in~\eqref{Eq:Cx8};
 then interchanging the order of summation over $w$
 and integration with respect to $\zeta$, we get
 \begin{equation}
  \label{Eq:Cxz6}
  \begin{split}
   & \CCSi^\chrg(t)=\left(\frac{x+1}2\right)^{1-\fst}
                         \frac 1{2\pi i}\oint_\gamma
       \left(\sum_{u=0}^\infty (-1)^u
              \tx^{a_u}
             \vphantom{%
                \frac 1{2\pi i}\oint_\Gamma
                \frac{(\zeta-1)^v
                        \left(\dfrac{\zeta+1}{\zeta-1}\right)^\fst}
                     {\zeta-z}
                                       \sum_{w=0}^\infty
              \left(h
                \frac{\zeta^2-1}
                     {\zeta-z}
              \right)^w d\zeta}
       \right.                                               \\
   & \qquad
       {}\times%
              (x+1)^{\beta_u}
        \frac{\tauxi^{|\alpha_u|+\beta_u}}
             {(\xi-1)^\fst(\xi-x)^{\beta_u-\fst+1}}
                      \frac{z+1}{2^{u+1}}                    \\
   & \quad
       {}\times%
                \frac 1{2\pi i}\oint_\Gamma
                \frac{(\zeta-1)^u
                        \left(\dfrac{\zeta-1}{\zeta+1}\right)^\fst}
                     {\zeta-z}
                                       \sum_{w=0}^\infty
              \left(h
                \frac{\zeta^2-1}
                     {\zeta-z}
              \right)^w d\zeta                               \\[2ex]
   & \qquad \qquad \qquad
           {}+\sum_{v=1}^\infty (-1)^v
              \tx^{a_v}
              (x+1)^{\beta_v}                                \\
   & \qquad
       {}\times%
        \frac{\tauxi^{|\alpha_v|+\beta_v}}
             {(\xi-1)^\fst(\xi-x)^{\beta_v-\fst+1}}
                      \left(\frac{z-1}{z+1}\right)^\fst
                      \frac{z+1}{2^{v+1}}                    \\
   &
       \left.\vphantom{%
           {}+\sum_{v=1}^\infty (-1)^v
              \tx^{a_v}
              (x+1)^{\beta_v}
        \frac{\tauxi^{|\alpha_v|+\beta_v}}
             {(\xi-1)^\fst(\xi-x)^{\beta_v-\fst+1}}}
       {}\times%
                \frac 1{2\pi i}\oint_\Gamma
                \frac{(\zeta-1)^v
                        \left(\dfrac{\zeta+1}{\zeta-1}\right)^\fst}
                     {\zeta-z}
                                       \sum_{w=0}^\infty
              \left(h
                \frac{\zeta^2-1}
                     {\zeta-z}
              \right)^w d\zeta
       \right) d\xi,
  \end{split}
 \end{equation}
 where
 \begin{equation}
  \label{Eq:h}
  h=\tx^{\rl1}\frac{z-1}4.
 \end{equation}

 Denote by $I_u$ and $I_v$ the inner integrals as follows:
 \begin{equation}
  \label{Eq:Iu1}
  I_u=\frac 1{2\pi i}\oint_\Gamma
             \frac{(\zeta-1)^u
                     \left(\dfrac{\zeta-1}{\zeta+1}\right)^\fst}
                  {\zeta-z}
                                    \sum_{w=0}^\infty
           \left(h
             \frac{\zeta^2-1}
                  {\zeta-z}
           \right)^w d\zeta
 \end{equation}
 and
 \begin{equation}
  \label{Eq:Iv1}
  I_v=\frac 1{2\pi i}\oint_\Gamma
             \frac{(\zeta-1)^v
                     \left(\dfrac{\zeta+1}{\zeta-1}\right)^\fst}
                  {\zeta-z}
                                    \sum_{w=0}^\infty
           \left(h
             \frac{\zeta^2-1}
                  {\zeta-z}
           \right)^w d\zeta.
 \end{equation}

 Note that for
 \[
  \left|h
    \frac{\zeta^2-1}
         {\zeta-z}
  \right| < 1,
 \]
 we have convergent series in these integrals.

 Consider~\eqref{Eq:Iu1}. We can write
 \begin{align*}
  I_u &= \frac 1{2\pi i}\oint_\Gamma
                \frac{(\zeta-1)^u
                        \left(\dfrac{\zeta-1}{\zeta+1}\right)^\fst}
                     {\zeta-z}
         \frac 1%
               {1-h \dfrac{\zeta^2-1}
                          {\zeta-z}} d\zeta \\
      &= \frac 1{2\pi i}\oint_\Gamma
                \frac{(\zeta-1)^u
                        \left(\dfrac{\zeta-1}{\zeta+1}\right)^\fst}
                     {\zeta-z-h(\zeta^2-1)} d\zeta.
 \end{align*}
 Consider the integrand
 \[
  \begin{split}
   \eta_u(\zeta) &= \frac{\chi_u(\zeta)}
                         {\psi_u(\zeta)} \\
                 &= \frac{(\zeta-1)^u
                            \left(\dfrac{\zeta-1}{\zeta+1}\right)^\fst}
                         {\zeta-z-h(\zeta^2-1)}.
  \end{split}
 \]
 We have denominator of the integrand
 $\psi_u(\zeta)=\left(\zeta-z-h(\zeta^2-1)\right)(\zeta+1)$
 which has three zeros:
 \begin{equation}
  \label{Eq:den_roots1}
  \zeta_{1,2}=\frac{1\mp\sqrt{1-4hz+4h^2}}{2h}
 \end{equation}
 and $\zeta_3=-1$,
 where $\displaystyle \lim_{h \to 0} \zeta_1=z$
 and $\zeta_2$ becomes infinite as $h \to 0$.

 Hence, for sufficiently small $h$, one may suppose that there is just
 one zero: $\zeta_1$ lying inside $\Gamma$, and the integrand has
 a simple pole inside the contour of integration with residue
 \[
  \Res_{\;\zeta=\zeta_1}(\eta_u(\zeta))
                 =\left.\frac{\chi_u (\zeta)}
                             {\psi_u'(\zeta)}\right|_{\zeta=\zeta_1},
 \]
 where $\psi_u'(\zeta)=1-2h\zeta$.

 Since $I_u=\Res_{\zeta=\zeta_1}(\eta_u(\zeta))$, then
 \begin{equation}
  \label{Eq:ResIu}
  I_u=\frac{\left(\dfrac{1-R}{2h}-1\right)^u
                   \left(\dfrac{1-2h-R}{1+2h-R}\right)^\fst}R,
 \end{equation}
 where
 \begin{equation}
  \label{Eq:R}
  R=\sqrt{1-4hz+4h^2}.
 \end{equation}

 Now, consider~\eqref{Eq:Iv1}. We can rewrite
 \[
  I_v=\frac 1{2\pi i}\oint_\Gamma
             \frac{(\zeta-1)^v
                     \left(\dfrac{\zeta+1}{\zeta-1}\right)^\fst}
                  {\zeta-z-h(\zeta^2-1)} d\zeta.
 \]
 Consider the integrand
 \[
  \begin{split}
   \eta_v(\zeta) &= \frac{\chi_v(\zeta)}
                         {\psi_v(\zeta)} \\
                 &= \frac{(\zeta-1)^v
                            \left(\dfrac{\zeta+1}{\zeta-1}\right)^\fst}
                         {\zeta-z-h(\zeta^2-1)}.
  \end{split}
 \]
 Recall that $v \geq 1$, then we have denominator of the integrand
 $\psi_v(\zeta)= \zeta-z-h(\zeta^2-1) $.
 In this case we also have~\eqref{Eq:den_roots1}
 as zeros of the denominator with the same properties.
 Then $I_v=\Res_{\zeta=\zeta_1}(\eta_v(\zeta))$ or
 \begin{equation}
  \label{Eq:ResIv}
  I_v=\frac{\left(\dfrac{1-R}{2h}-1\right)^v
                   \left(\dfrac{1+2h-R}{1-2h-R}\right)^\fst}R.
 \end{equation}
 By substituting~\eqref{Eq:ResIu} and~\eqref{Eq:ResIv}
 to~\eqref{Eq:Cxz6} we have
 \begin{equation}
  \label{Eq:ResCxz4}
  \begin{split}
   & \CCSi^\chrg(t)=\left(\frac{x+1}2\right)^{1-\fst}                 \\
   & \qquad
       {}\times
                   \frac 1{2\pi i}\oint_\gamma
       \Biggl(\frac 1{\xi-1}\left(\frac{\xi-x}{\xi-1}\right)^\fst
              \frac{z+1}{2R}
              \left(\frac{1-2h-R}{1+2h-R}\right)^\fst                 \\
   & \quad
       {}\times
              \sum_{u=0}^\infty (-1)^u
               \tx^{a_u}
               \left(\frac{x  +1}{\xi-x}\right)^{\beta_u}
               \tauxi^{|\alpha_u|+\beta_u}
                  \left(\frac{1-2h-R}{4h}\right)^u                    \\
   & \qquad
           {}+\frac 1{\xi-1}\left(\frac{\xi-x}{\xi-1}\right)^\fst
              \frac{z+1}{2R}
              \left(\frac{z-1}{z+1}\frac{1+2h-R}{1-2h-R}\right)^\fst  \\
   &
       {}\times
              \sum_{v=1}^\infty (-1)^v
               \tx^{a_v}
               \left(\frac{x  +1}{\xi-x}\right)^{\beta_v}
               \tauxi^{|\alpha_v|+\beta_v}
                  \left(\frac{1-2h-R}{4h}\right)^v
       \Biggr) d\xi.
  \end{split}
 \end{equation}

 Recall~\eqref{Eq:alphau}, \eqref{Eq:au}, \eqref{Eq:betau},
 \eqref{Eq:alphav}, \eqref{Eq:av}, and~\eqref{Eq:betav} (see Table~\ref{Tab:DiagSum}),
 then according to $\alpha_u$ and $\alpha_v$
 properties, we need to break these series on two parts each;
 one finite and one infinite series. Define these series as follows:
 \begin{align*}
  \sum_{u=0}^\infty (\cdot)^u & =   \sum_{u=0}^{\ufs}     (\cdot)^u
                                  + \sum_{u=\uis}^\infty  (\cdot)^u, \\
  \sum_{v=1}^\infty (\cdot)^u & =   \sum_{v=1}^{\vfsc}    (\cdot)^v
                                  + \sum_{v=\visc}^\infty (\cdot)^v.
 \end{align*}

 \renewcommand{\cts}[1]{\tilde \sigma} 
 By $\cts1$ denote the shifted charge; then
 by $\uvc$  denote the quotient of $\cts1$ and $\rl1$
 \begin{align*}
  \cts1 &= \chrg-\ats+\bts, \notag \\
  \uvc  &= \left\lceil \frac{\cts1}
                            {\rl1}
           \right\rceil.
 \end{align*}

 \renewcommand{\ufs}{ \uvc - 1 }
 \renewcommand{\uis}{ \max \left(
                            \uvc , 0
                           \right)}
 \renewcommand{\visc}{\max \left(
                            -\uvc + 1 , 1
                           \right)}
 \renewcommand{\vfsc}{ -\uvc }
 Consider~\eqref{Eq:ResCxz4} as four sums of integrals along the contour $\gamma$.
 Each of these integrals includes one of the series above.
 Let $J_1$, $J_2$, $J_3$, and $J_4$ denote these integrals.
 \begin{align*}
  J_1 &=  \frac 1{2\pi i}\oint_\gamma
             \Psi_1(\xi) d\xi,        &
  J_2 &=  \frac 1{2\pi i}\oint_\gamma
             \Psi_2(\xi) d\xi,        \\
  J_3 &= -\frac 1{2\pi i}\oint_\gamma
             \Psi_3(\xi) d\xi,        &
  J_4 &= -\frac 1{2\pi i}\oint_\gamma
             \Psi_4(\xi) d\xi,
 \end{align*}
 where by $\Psi_1$, $\Psi_2$, $\Psi_3$, and $\Psi_4$
 we denote the integrands of $J_1$, $J_2$, $J_3$, and $J_4$.
 \begin{equation*}
   \Psi_1(\xi) =
            \frac{\tauxi^{ \cts1}}
                 {\xi-x}\left(\frac{\xi-x}{\xi-1}\right)^\fst
            \frac{z+1}{2R}
            \left(\frac{1-2h-R}{1+2h-R}\right)^\fst
                \sum_{u=0}^{\ufs}
                 T_1^u
   ,
 \end{equation*}
 \begin{equation}
  \label{Eq:J2}
  \begin{split}
   \Psi_2(\xi) &= 
            \frac{\tauxi^{-\cts2}}
                 {(\xi-x)(\xi+1)^\fst}
            \frac{z+1}{2R}
            \left(\frac{1-2h-R}{1+2h-R}\right)^\fst                \\
               &{}\quad\times
                 \sum_{u=\uis}^\infty
                  T_2^u
   ,
  \end{split}
 \end{equation}
 \begin{equation*}
  \begin{split}
   \Psi_3(\xi) &= 
            \frac{\tauxi^{-\cts2}}
                 {(\xi-x)(\xi+1)^\fst}
            \frac{z+1}{2R}
            \left(\frac{z-1}{z+1}\frac{1+2h-R}{1-2h-R}\right)^\fst \\
               &{}\quad\times
                \sum_{v=1}^{\vfsc}
                 T_1^v
   ,
  \end{split}
 \end{equation*}
 \begin{equation}
  \label{Eq:J4}
  \begin{split}
   \Psi_4(\xi) &=
            \frac{\tauxi^{ \cts1}}
                 {\xi-x}\left(\frac{\xi-x}{\xi-1}\right)^\fst
            \frac{z+1}{2R}
            \left(\frac{z-1}{z+1}\frac{1+2h-R}{1-2h-R}\right)^\fst \\
          &{}\quad\times
                 \sum_{v=\visc}^\infty
                  T_2^v
   ,
  \end{split}
 \end{equation}
 where
 \begin{align}
  \label{Eq:T1}
  T_1 &= -\left(\tx/\tauxi\right)^{\rl1}
                   \frac{1-2h-R}{4h}, \\
  \label{Eq:T2}
  T_2 &= -\tauxi^{\rl1}
                   \frac{1-2h-R}{4h};
 \end{align}
 then~\eqref{Eq:ResCxz4} becomes
 \begin{equation}
  \label{Eq:gfCt}
  \begin{split}
   \CCSi^\chrg(t)=
      \frac{       t^{\fmc+\ats}}
           {(1-t)^{1-\fst}}
   & \left( J_{1}+J_{4} \right) \\
   &
   {}+\frac{2^\fst t^{\fpc+\bts}}
           {1-t}
     \left( J_{2}+J_{3} \right).
  \end{split}
 \end{equation}
 Note that for
 \[
  \left| T_2 \right| < 1,
 \]
 we have the convergent geometric series
 in~\eqref{Eq:J2} and~\eqref{Eq:J4}.
 Therefore we can rewrite these integrands in closed form.
 Then we substitute~\eqref{Eq:h} for $h$;
 then, in turn, $\frac{1+s}{1-s}$ for $z$,
 and~\eqref{Eq:s} for $s$.
 Now, for the sake of simplicity
 and for eliminating, where possible, the imaginary parts of the denominators,
 we perform straightforward calculations and obtain
 \newcommand{\Pxi}[1]{P_{#1}}
 \newcommand{\Q}{Q}
 \begin{equation}
  \label{Eq:IntegrandJ1}
  \begin{split}
   \Psi_1(\xi)=
            \frac{\tau^{ \cts1}}{2\Rt}
          \left(\bd(\xi+1)\right)^\fst
        & \frac{ H_1
                +\left(1-\fst\tau^{\rl1}\right) \Rt}
               {\Q}                                           \\
        & \qquad
          {}\times\left(
                   1-T_1^{\uex}
                  \right),
  \end{split}
 \end{equation}
 \begin{equation}
  \label{Eq:IntegrandJ2}
  \begin{split}
   \Psi_2(\xi)=
            \frac{\tau^{-\cts2}}{2\Rt}
          \left(\frac{\bd(\xi-1)}
                     {\xi-x}
          \right)^\fst
        & \frac{ H_2
                +\left(1-\fst\left(t/\tau\right)^{\rl1}
                             \right)            \Rt}
               {\Q}                                           \\
        & \qquad \qquad
          {}\times%
                     T_2^{\uex},
  \end{split}
 \end{equation}
 \begin{equation}
  \label{Eq:IntegrandJ3}
  \begin{split}
   \Psi_3(\xi)=
            \frac{\tau^{-\cts2}}{2\Rt}
          \left(\frac{\bd(\xi-1)}
                     {\xi-x}
          \right)^\fst
        & \frac{ H_2
                -\left(1-\fst\left(t/\tau\right)^{\rl1}
                             \right)            \Rt}
               {\Q}                                           \\
        &
          {}\times\left(
                   1-T_1^{\vexa}
                  \right),
  \end{split}
 \end{equation}
 \begin{equation}
  \label{Eq:IntegrandJ4}
  \begin{split}
   \Psi_4(\xi)=
            \frac{\tau^{ \cts1}}{2\Rt}
          \left(\bd(\xi+1)\right)^\fst
        & \frac{ H_1
                -\left(1-\fst\tau^{\rl1}\right)  \Rt}
               {\Q}                                           \\
        & \qquad \qquad \quad
          {}\times%
                     T_2^{\vexa},
  \end{split}
 \end{equation}
 where
 \begin{align*}
  H_1 &= \fst \Pxi1+\tau^{\fst\rl2} \Pxi3,             \\
  H_2 &= \fst \left( t/\tau\right )^{\rl1} \Pxi2+\Pxi4;
 \end{align*}
 $\bk$ is then defined like~\eqref{Eq:b} as follows:
 \[
  \bk=\frac{t^{k+2}}
           {2(1-t)}.
 \]
 Now~\eqref{Eq:T1} and~\eqref{Eq:T2} become
 \begin{align}
  \label{Eq:T1a}
  T_1 &= -\left(t/\tau\right)^{\rl1}
                 \dfrac{P_0-\Rt}
                       {2\bk\left(\xi^2-1\right)}, \\
  \label{Eq:T2a}
  T_2 &= -\tau^{\rl1}
                 \dfrac{P_0-\Rt}
                       {2\bk(\xi^2-1)}.
 \end{align}

\begin{figure}[!t]
  \begin{center}
  \unitlength=0.8mm
  \begin{picture}(75,50)
   \put(0,25){\vector(1,0){75}}          
   \put(70,26.5){$\Re$}                  
   \put(30,25){\circle*{1}}
   \put(22,26.5){$\xi_5=x$}
   \qbezier(55,    25   )(55,   31.7)(51.65, 37.5 ) 
   \qbezier(51.65, 37.5 )(48.3, 43.3)(42.5,  46.65) 
   \qbezier(42.5,  46.65)(36.7, 50  )(30,    50   ) 
   \qbezier(30,    50   )(23.3, 50  )(17.5,  46.65) 
   \qbezier(17.5,  46.65)(11.7, 43.3)( 8.35, 37.5 ) 
   \qbezier( 8.35, 37.5 )( 5,   31.7)( 5,    25   ) 
   \qbezier( 5,    25   )( 5,   18.3)( 8.35, 12.5 ) 
   \qbezier( 8.35, 12.5 )(11.7,  6.7)(17.5,   3.35) 
   \qbezier(17.5,   3.35)(23.3,  0  )(30,     0   ) 
   \qbezier(30,     0   )(36.7,  0  )(42.5,   3.35) 
   \qbezier(42.5,   3.35)(48.3,  6.7)(51.65, 12.5 ) 
   \qbezier(51.65, 12.5 )(55,   18.3)(55,    25   ) 
   \put(42.5, 46.7 ){\vector(-2, 1){1}}
   \put(17.5, 46.6 ){\vector(-2,-1){1}}
   \put(17.5,  3.35){\vector( 2,-1){1}}
   \put(42.5,  3.35){\vector( 2, 1){1}}
   \put(35,25){\circle*{1}}
   \put(36,26.5){$\xi_3$}
   \put(50,25){\circle*{1}}
   \put(51,26.5){$\xi_1$}
   \put(45, 46){$\gamma$}                
   {\linethickness{0.75mm}               
    \put(35,25){\line(1,0){15}}}
  \end{picture}
  \end{center}
  \caption{Contour $\gamma$ which is the circle
           (or any Jordan curve) about $x$}
  \label{Fig:CircleContour}
 \end{figure}
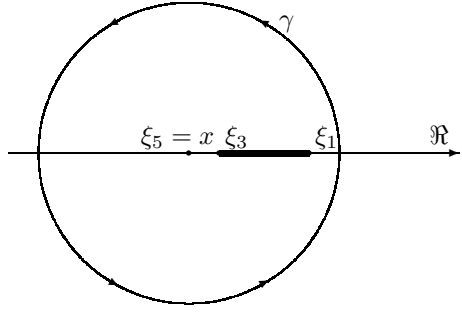

 \renewcommand{\Pxi}[1]{\ifthenelse{\equal{#1}{0}}%
                       {\xi-x-\left( \bd+\bk \right) \left( \xi^2-1 \right)}{}
                        \ifthenelse{\equal{#1}{1}}%
                       {\xi-x-\left( \bd-\bk \right) \left( \xi^2-1 \right)}{}
                        \ifthenelse{\equal{#1}{2}}%
                       {\xi-x-\left( \bk-\bd \right) \left( \xi^2-1 \right)}{}
                        \ifthenelse{\equal{#1}{3}}%
                       {\xi-x-\left( \bd\left( 1-2\tau^{(1-\fst)\rl2}%
                                        \right)%
                                    +\bk%
                              \right) \left( \xi^2-1 \right)}{}
                        \ifthenelse{\equal{#1}{4}}%
                       {\xi-x-\left( \bd%
                                    +\bk\left( 1-2\tau^{(\fst-1)\rl2}%
                                        \right)%
                              \right) \left( \xi^2-1 \right)}{}}
 \renewcommand{\Q}{\xi-x-\left( \bd \left( 1-\tau^{ \rl1} \right)
                               +\bk \left( 1-\tau^{-\rl2} \right)
                         \right) \left( \xi^2-1 \right)}
 $P_0, \dots, P_4$
 are polynomials of degree $2$ in the complex variable $\xi$.
 \begin{align*}
  P_0 &= \Pxi0, \\
  P_1 &= \Pxi1, \\
  P_2 &= \Pxi2, \\
  P_3 &= \Pxi3, \\
  P_4 &= \Pxi4.
 \end{align*}
 Denominator functions  $Q$, and $\Rt$ are defined as
 \begin{align}
  \label{Eq:Q}
  Q   &= \Q,                                                       \\
  \label{Eq:Rt}
  \Rt &= \pm \sqrt{P_0^2
                           -4\bd\bk\left( \xi^2-1 \right)^2}.
 \end{align}
 The functions~\eqref{Eq:Q} and~\eqref{Eq:Rt}
 define singularities of integrands~%
 \eqref{Eq:IntegrandJ1}, \eqref{Eq:IntegrandJ2}, \eqref{Eq:IntegrandJ3},
 and~\eqref{Eq:IntegrandJ4}.
 Indeed, $\Rt$ have 4 roots; these all roots are real:
 \begin{align*}
  \xi_{1,2} &=\frac{1\mp\sqrt{1-4\left(\sqrt{\bd}+\sqrt{\bk}\right)^2x%
                               +4\left(\sqrt{\bd}+\sqrt{\bk}\right)^4}}
                   {2\left(\sqrt{\bd}+\sqrt{\bk}\right)^2},             \\
  \xi_{3,4} &=\frac{1\mp\sqrt{1-4\left(\sqrt{\bd}-\sqrt{\bk}\right)^2x%
                               +4\left(\sqrt{\bd}-\sqrt{\bk}\right)^4}}
                   {2\left(\sqrt{\bd}-\sqrt{\bk}\right)^2}.
 \end{align*}
 Note, that~\eqref{Eq:Rt} becomes imaginary on
 $[\xi_3, \xi_1]$ and $[\xi_2, \xi_4]$;
 therefore~\eqref{Eq:IntegrandJ1},~\dots, \eqref{Eq:IntegrandJ4}
 have branch cuts on these segments.
 Obviously, that $Q=0$ if $\xi=x$.
 Therefore, there might be a singularity at point
 $\xi_5=x$ for integrands~%
 \eqref{Eq:IntegrandJ1},~\dots, \eqref{Eq:IntegrandJ4}.
 It may be simply proven that $\xi_1$, $\xi_3$, and $\xi_5$
 lie inside the contour $\gamma$
 as shown on Fig.~\ref{Fig:CircleContour}.
 Thick line inside the contour denotes the branch cut
 on $[\xi_3, \xi_1]$.

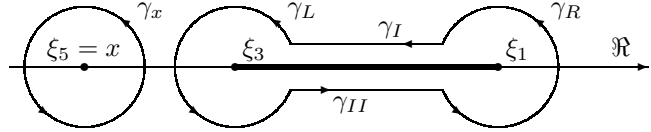
\begin{figure}[!t]
  \begin{center}
  \unitlength=1mm
  \begin{picture}(85,25)
   \put(0,12.5){\vector(1,0){85}}        
   \put(80,14.0){$\Re$}                  
   \put(10,12.5){\circle*{1}}
   \put(5,14.0){$\xi_5=x$}
   \qbezier(18.00, 12.50)(18.00, 14.09)(17.39, 15.56)
   \qbezier(17.39, 15.56)(16.78, 17.03)(15.66, 18.16)
   \qbezier(15.66, 18.16)(14.53, 19.28)(13.06, 19.89)
   \qbezier(13.06, 19.89)(11.59, 20.50)(10.00, 20.50)
   \qbezier(10.00, 20.50)( 8.41, 20.50)( 6.94, 19.89)
   \qbezier( 6.94, 19.89)( 5.47, 19.28)( 4.34, 18.16)
   \qbezier( 4.34, 18.16)( 3.22, 17.03)( 2.61, 15.56)
   \qbezier( 2.61, 15.56)( 2.00, 14.09)( 2.00, 12.50)
   \qbezier( 2.00, 12.50)( 2.00, 10.91)( 2.61,  9.44)
   \qbezier( 2.61,  9.44)( 3.22,  7.97)( 4.34,  6.84)
   \qbezier( 4.34,  6.84)( 5.47,  5.72)( 6.94,  5.11)
   \qbezier( 6.94,  5.11)( 8.41,  4.50)(10.00,  4.50)
   \qbezier(10.00,  4.50)(11.59,  4.50)(13.06,  5.11)
   \qbezier(13.06,  5.11)(14.53,  5.72)(15.66,  6.84)
   \qbezier(15.66,  6.84)(16.78,  7.97)(17.39,  9.44)
   \qbezier(17.39,  9.44)(18.00, 10.91)(18.00, 12.50)
   \put(15.66, 18.30){\vector(-3, 2){1}}
   \put( 4.34,  6.84){\vector( 3,-2){1}}
   \put(17, 19.5){$\gamma_x$}            
   \put(30,12.5){\circle*{1}}
   \put(31,14.0){$\xi_3$}
   \qbezier(37.39, 15.56)(36.78, 17.03)(35.66, 18.16)
   \qbezier(35.66, 18.16)(34.53, 19.28)(33.06, 19.89)
   \qbezier(33.06, 19.89)(31.59, 20.50)(30.00, 20.50)
   \qbezier(30.00, 20.50)(28.41, 20.50)(26.94, 19.89)
   \qbezier(26.94, 19.89)(25.47, 19.28)(24.34, 18.16)
   \qbezier(24.34, 18.16)(23.22, 17.03)(22.61, 15.56)
   \qbezier(22.61, 15.56)(22.00, 14.09)(22.00, 12.50)
   \qbezier(22.00, 12.50)(22.00, 10.91)(22.61,  9.44)
   \qbezier(22.61,  9.44)(23.22,  7.97)(24.34,  6.84)
   \qbezier(24.34,  6.84)(25.47,  5.72)(26.94,  5.11)
   \qbezier(26.94,  5.11)(28.41,  4.50)(30.00,  4.50)
   \qbezier(30.00,  4.50)(31.59,  4.50)(33.06,  5.11)
   \qbezier(33.06,  5.11)(34.53,  5.72)(35.66,  6.84)
   \qbezier(35.66,  6.84)(36.78,  7.97)(37.39,  9.44)
   \put(35.66, 18.30){\vector(-3, 2){1}}
   \put(24.34,  6.84){\vector( 3,-2){1}}
   \put(65,12.5){\circle*{1}}
   \put(66,14.0){$\xi_1$}
   \qbezier(73.00, 12.50)(73.00, 14.09)(72.39, 15.56)
   \qbezier(72.39, 15.56)(71.78, 17.03)(70.66, 18.16)
   \qbezier(70.66, 18.16)(69.53, 19.28)(68.06, 19.89)
   \qbezier(68.06, 19.89)(66.59, 20.50)(65.00, 20.50)
   \qbezier(65.00, 20.50)(63.41, 20.50)(61.94, 19.89)
   \qbezier(61.94, 19.89)(60.47, 19.28)(59.34, 18.16)
   \qbezier(59.34, 18.16)(58.22, 17.03)(57.61, 15.56)
   \qbezier(57.61,  9.44)(58.22,  7.97)(59.34,  6.84)
   \qbezier(59.34,  6.84)(60.47,  5.72)(61.94,  5.11)
   \qbezier(61.94,  5.11)(63.41,  4.50)(65.00,  4.50)
   \qbezier(65.00,  4.50)(66.59,  4.50)(68.06,  5.11)
   \qbezier(68.06,  5.11)(69.53,  5.72)(70.66,  6.84)
   \qbezier(70.66,  6.84)(71.78,  7.97)(72.39,  9.44)
   \qbezier(72.39,  9.44)(73.00, 10.91)(73.00, 12.50)
   \put(70.66, 18.30){\vector(-3, 2){1}}
   \put(59.34,  6.84){\vector( 3,-2){1}}
   \put(37.39, 15.56){\line(1,0){20.22}}
   \put(37.39,  9.44){\line(1,0){20.22}}
   \put(53.16, 15.56){\vector(-1,0){1}}
   \put(41.84,  9.44){\vector( 1,0){1}}
   \put(37, 19.5){$\gamma_L$}            
   \put(72, 19.5){$\gamma_R$}
   \put(49.0, 17.0){$\gamma_I$}
   \put(43.0,  7.0){$\gamma_{II}$}
   {\linethickness{0.75mm}               
    \put(30,12.5){\line(1,0){35}}}
  \end{picture}
  \end{center}
  \caption{Dumbbell-shaped contour about the branch cut on $[\xi_3, \xi_1]$
           and a circle contour about $\xi_5$.}
  \label{Fig:DumbbellContour}
 \end{figure}

 Instead of the contour $\gamma$,
 we can consider two contours;
 the first dumbbell-shaped contour whose ``balls'' contain,
 respectively, $\xi_3$ and $\xi_1$
 and the second contour $\gamma_x$
 as shown on Fig.~\ref{Fig:DumbbellContour}.
 In this case we can write $J_1$, say, by the following
 \begin{equation}
  \label{Eq:J1DMBcontour}
  \begin{split}
   J_1 &= \frac{1}{2\pi i}\left(\oint_{\gamma_L}    \Psi_1(\xi)d\xi
                               +\oint_{\gamma_I}    \Psi_1(\xi)d\xi
                                \vphantom{
                               +\oint_{\gamma_R}    \Psi_1(\xi)d\xi
                               +\oint_{\gamma_{II}} \Psi_1(\xi)d\xi
                               +\oint_{\gamma_x}    \Psi_1(\xi)d\xi}
                          \right.                                    \\
       &
                          \left.\vphantom{
                                \oint_{\gamma_L}    \Psi_1(\xi)d\xi
                               +\oint_{\gamma_I}    \Psi_1(\xi)d\xi}
                               +\oint_{\gamma_R}    \Psi_1(\xi)d\xi
                               +\oint_{\gamma_{II}} \Psi_1(\xi)d\xi
                             {}+\oint_{\gamma_x}    \Psi_1(\xi)d\xi
                          \right).
  \end{split}
 \end{equation}
 We can show that the integrals along the two circles
 $\gamma_L$ and $\gamma_L$ vanish in the limit as the
 radius $\rDMB$ of these circles tends to $0$.
 Indeed, we can rewrite~\eqref{Eq:Rt} as
 \begin{equation*}
  \Rt=(\bd-\bk)\sqrt{(\xi-\xi_1)(\xi-\xi_2)(\xi-\xi_3)(\xi-\xi_4)}.
 \end{equation*}
 Therefore, for arbitrarily small $\rDMB$, we have
 \begin{align}
  \label{Eq:RtRootxi1}
  \begin{split}
   \left.\Rt\right|_{|\xi-\xi_1| \leq \rDMB}
   &=    (\bd-\bk)\sqrt{(\xi-\xi_2)(\xi-\xi_3)(\xi-\xi_4)} \\
   &      \qquad {}\times \sqrt{(\xi-\xi_1)},
  \end{split}                                            \\
  \label{Eq:RtRootxi3}
  \begin{split}
  \left.\Rt\right|_{|\xi-\xi_3| \leq \rDMB}
   &= \pm(\bd-\bk)\sqrt{(\xi-\xi_1)(\xi-\xi_2)(\xi-\xi_4)} \\
   &      \qquad {}\times \sqrt{(\xi-\xi_3)}.
  \end{split}
 \end{align}
 Now, we estimate integrals $\oint_{\gamma_L} \Psi_1(\xi)d\xi$
                        and $\oint_{\gamma_R} \Psi_1(\xi)d\xi$ as follows.
 \renewcommand{\Pxi}[1]{P_{#1}}
 \renewcommand{\Q}{Q}
 First, substituting~\eqref{Eq:RtRootxi1} and~\eqref{Eq:RtRootxi3}
 for $\Rt$ in~\eqref{Eq:IntegrandJ1}, we see that
 \begin{equation*}
  \begin{split}
   &        \frac{\tau^{ \cts1}}
                 {2(\bd-\bk)\sqrt{(\xi-\xi_2)(\xi-\xi_3)(\xi-\xi_4)}}
          \left(\bd(\xi+1)\right)^\fst                                \\
   & \qquad \qquad
          {}\times%
          \frac{ H_1
                +\left(1-\fst\tau^{\rl1}\right) \Rt}
               {\Q}
                  \left(
                   1-T_1^{\uex}
                  \right)
   \leq M
  \end{split}
 \end{equation*}
 are bounded above at the point $\xi_3$ and it's nearest neighborhood
 as well as
 \begin{equation*}
  \begin{split}
   &        \frac{\tau^{ \cts1}}
                 {2(\bd-\bk)\sqrt{(\xi-\xi_1)(\xi-\xi_2)(\xi-\xi_4)}}
          \left(\bd(\xi+1)\right)^\fst                                \\
   & \qquad \qquad
          {}\times%
          \frac{ H_1
                +\left(1-\fst\tau^{\rl1}\right) \Rt}
               {\Q}
                  \left(
                   1-T_1^{\uex}
                  \right)
   \leq M
  \end{split}
 \end{equation*}
 are bounded above at the point $\xi_1$ and it's nearest neighborhood.
 Then we show that
 \begin{align*}
  \oint_{\gamma_R} \Psi_1(\xi)d\xi
   & \leq M\oint_{\gamma_R} \frac{d\xi}{\sqrt{\xi-\xi_1}}, \\
  \oint_{\gamma_L} \Psi_1(\xi)d\xi
   & \leq M\oint_{\gamma_L} \frac{d\xi}{\sqrt{\xi-\xi_3}}.
 \end{align*}
 Using parameterization of the contours
 $\gamma_R$ by $\xi(\vartheta)=\xi_1+\rDMB e^{i\vartheta}$,
 $d\xi=i\rDMB e^{i\vartheta}d\vartheta$
 with $\vartheta\in[-\pi, \pi]$ and
 $\gamma_L$ by $\xi(\vartheta)=\xi_3+\rDMB e^{i\vartheta}$,
 $d\xi=i\rDMB e^{i\vartheta}d\vartheta$
 with $\vartheta\in[0, 2\pi]$,
 we have
 \begin{align*}
  \oint_{\gamma_R} \frac{d\xi}{\sqrt{\xi-\xi_1}}
   &=  4i\sqrt\rDMB                              \\
  \oint_{\gamma_L} \frac{d\xi}{\sqrt{\xi-\xi_3}}
   &= -4\sqrt\rDMB
 \end{align*}
 and we can see that
 $\oint_{\gamma_R}    \Psi_1(\xi)d\xi$ and
 $\oint_{\gamma_L}    \Psi_1(\xi)d\xi$
 vanish as $\rDMB$ tends to zero.

 Hence, there remain two integrals
 $\oint_{\gamma_I}    \Psi_1(\xi)d\xi$ and
 $\oint_{\gamma_{II}} \Psi_1(\xi)d\xi$
 along the line segments just above and just below the cut,
 and one integral
 $\oint_{\gamma_x}    \Psi_1(\xi)d\xi$
 (see Fig.~\ref{Fig:DumbbellContour}).

 Along the upper and along the lower segments, we can rewrite
 \begin{align*}
       \oint_{\gamma_I}  & \Psi_1(\xi) d\xi
     + \oint_{\gamma_{II}} \Psi_1(\xi) d\xi             \\
   &=  \int_{\xi_3 + \rDMB + i 0}^{\xi_1 - \rDMB + i 0}
                           \Psi_1(\xi) d\xi
     + \int_{\xi_1 - \rDMB - i 0}^{\xi_3 + \rDMB - i 0}
                           \Psi_1(\xi) d\xi             \\
   &=  i \int_{\xi_3}^{\xi_1}
                 \Im\left( \Psi_1(\xi) \right) d\xi
     - i \int_{\xi_1}^{\xi_3}
                 \Im\left( \Psi_1(\xi) \right) d\xi,
   \quad \text{as} \; \rDMB \to 0.
 \end{align*}
 This directly follows from the fact that $\Rt$ is the only function
 which becomes imaginary into these limits of integration.
 Obviously, that~\eqref{Eq:J1DMBcontour} becomes
 \begin{equation}
  \label{Eq:J1seg}
  J_1    =  \frac{1}{\pi}\int_{\xi_3}^{\xi_1}
                 \Im\left( \Psi_1(\xi) \right) d\xi
                              +\Res_{\xi=x}(\Psi_1(\xi)).
 \end{equation}
 In the same way, we get $J_2, \dots, J_4$
 in the form identical with~\eqref{Eq:J1seg}.

 It is not hard to show that
 \begin{equation}
  \label{Eq:T}
  \frac{\Pxi0-\Rt}
       {2\bk(\xi^2-1)}
       = t^{-\rl2/2} e^{i\varphi}
 \end{equation}
 between the limits of integration
 $[\xi_3, \xi_1]$, where we define $\varphi$ as
 \begin{align}
  \label{Eq:phi1}
  \varphi &= -i \ln   \left( \frac{P_0-\Rt}
                                  {2\sqrt{\bd\bk}(\xi^2-1)}
                      \right)                               \\
          &= -\arccos \left( \frac{P_0}
                                  {2\sqrt{\bd\bk}(\xi^2-1)}
                      \right).                              \notag
 \end{align}
 Equation~\eqref{Eq:phi1} directly follows from the fact
 that $\Rt$ becomes imaginary on segment $[\xi_3, \xi_1]$.

 \newcommand{\nT}{\hat n}	
 Now, recall~\eqref{Eq:T1a} and~\eqref{Eq:T2a};
 then consider power functions $T_1^{\nT}$ and $T_2^{\nT}$.
 Taken into account~\eqref{Eq:T},
 we can rewrite these functions as
 \begin{align*}
  T_1^{\nT} &= (-1)^{\nT}\left(\frac{\sqrt{t}} \tau\right)^{ \nT \rl2}
                         \left(\cos{\nT \varphi}+i\sin{\nT \varphi}\right), \\
  T_2^{\nT} &= (-1)^{\nT}\left(\frac{\sqrt{t}} \tau\right)^{-\nT \rl2}
                         \left(\cos{\nT \varphi}+i\sin{\nT \varphi}\right),
 \end{align*}
 where $\nT$ denotes either $\uex$ or $\vexa$.

 Since we need only imaginary parts of~%
 \eqref{Eq:IntegrandJ1}, \eqref{Eq:IntegrandJ2}, \eqref{Eq:IntegrandJ3},
 and~\eqref{Eq:IntegrandJ4},
 then we can rewrite the sums $J_1+J_4$ and $J_2+J_3$
 in~\eqref{Eq:gfCt} as follows:
 \begin{equation}
  \label{Eq:J14ImFin}
  \begin{split}
   & J_1  + J_4
          = \frac{(-1)^{\uvc}}{2 \pi} \int_{\xi_3}^{\xi_1}
              \frac{\tau^{ \cts1}}{Q}
            \left(\bd(\xi+1)\right)^\fst                      \\
   & \quad
            {}\times%
                 \left(\frac{\sqrt{t}}\tau\right)^{ \uvc\rl2}
            \left(
              \frac{H_1}{|\Rt|}
                  \cos\left(\uvcd \varphi\right)
             -\left(1-\fst\tau^{\rl1}\right)
                  \sin\left(\uvcd \varphi\right)
            \right)
                                        d\xi                  \\
   & \qquad \qquad \qquad \qquad \qquad \qquad
         {}+\Res_{\xi=x}(\Psi_1(\xi))-\Res_{\xi=x}(\Psi_4(\xi))
  \end{split}
 \end{equation}
 and
 \begin{equation}
  \label{Eq:J23ImFin}
  \begin{split}
   & J_2  + J_3
          =-\frac{(-1)^{\uvc}}{2 \pi} \int_{\xi_3}^{\xi_1}
              \frac{\tau^{-\cts2}}{Q}
            \left(\frac{\bd(\xi-1)}
                       {\xi-x}
            \right)^\fst                                      \\
   &
            {}\times%
                 \left(\frac{\sqrt{t}}\tau\right)^{-\uvc\rl2}
            \left(
              \frac{H_2}{|\Rt|}
                  \cos\left(\uvcd \varphi\right)
             -\left(1-\fst\left(t/\tau\right)^{\rl1}\right)
                  \sin\left(\uvcd \varphi\right)
            \right)
                                        d\xi                  \\
   & \qquad \qquad \qquad \qquad \qquad \qquad
         {}+\Res_{\xi=x}(\Psi_2(\xi))-\Res_{\xi=x}(\Psi_3(\xi)).
  \end{split}
 \end{equation}

 To conclude the derivation of our generating function,
 it remains to note that
 \begin{align}
  \Res_{\xi=x}(\Psi_1(\xi))
    &= 0,                              \notag \\
  \label{Eq:ResPsi2}
  \Res_{\xi=x}(\Psi_2(\xi))
    &= \frac{\fst t^{\rl2 \uex -\cts2} (t-1)}
            {2\left( 1-t^{\rl1} \right)},     \\
  \Res_{\xi=x}(\Psi_3(\xi))
    &= 0,                              \notag \\
  \Res_{\xi=x}(\Psi_4(\xi))
    &= 0.                              \notag
 \end{align}
 The proof of~\eqref{Eq:ResPsi2} is straightforward but tedious and omitted here.

 Degree of radicand of $\Rt$ and properties of its roots
 define~\eqref{Eq:J14ImFin} and~\eqref{Eq:J23ImFin} as elliptic.
 Therefore, generating function~\eqref{Eq:Ct1} does not have a closed form.

 Thus, we just proved the following theorem.
 \begin{theorem}
  A recursion relation \eqref{Eq:Cc} does not have a closed form.
 \end{theorem}

 Note that from the residue theorem it follows that the limits $\xi_3$ and $\xi_1$
 may be replaced by $\xi_2$ and $\xi_4$ respectively
 with changing signs of~\eqref{Eq:J14ImFin}
 and~\eqref{Eq:J23ImFin}.

 \subsection{Generating Function of Sequence %
  $ \CCS_n^{ -n } , \CCS_n^{ -n + 2 } , \dots , \CCS_n^n $}

 Consider a finite power series
 \begin{equation*}
  \CCS_n(y)=\sum_{\substack{\chrg=-n \\ \chrg\ \mathrm{even}}}^{n}
                                     \CCS_n^\chrg y^{\chrg/2},
  \quad \text{or} \quad
  \CCS_n(y)=\sum_{\substack{\chrg=-n \\ \chrg\ \mathrm{odd}}}^{n}
                                     \CCS_n^\chrg y^{(\chrg-1)/2},
 \end{equation*}
 where $y \in \mathbb R$.

 Here we can also consider~\eqref{Eq:CdkrI} instead of $ \CCS_n^\chrg $.
 The calculations that let us got~\eqref{Eq:Ct5} yield now
 \begin{equation}
  \label{Eq:Cy1}
  \begin{split}
   \CCSi_n(y)=\sum_{ \substack{ m = \eo          \\
                                m\ \mathrm{even} \\
                                \text{or}        \\
                                m\ \mathrm{odd}} }^{ 2 n + \eo }
   & \left( \sum_{ u = 0 }^{ \tfrac{ m - \eo }2 - \fst }
            \sum_{ v = 0 }^{ \tfrac{ m - \eo }2 }        ( -1 )^{ u + v }
            \vphantom{%
      \binom{ \frac{ m - \eo }2 - \fst }u
      \binom{ \frac{ m - \eo }2        }v
      g(y)}
     \right.                                                              \\
   & \qquad
     \left. \vphantom{%
            \sum_{ u = 0 }^{ \tfrac{ m - \eo }2 - \fst }
            \sum_{ v = 0 }^{ \tfrac{ m - \eo }2 }        ( -1 )^{ u + v }}
      {}\times%
      \binom{ \frac{ m - \eo }2 - \fst }u
      \binom{ \frac{ m - \eo }2        }v
      g(y)
     \right),
  \end{split}
 \end{equation}
 where by $ g( y ) $ we denote the inner sum as follows:
 \begin{align}
  \label{Eq:gy1}
  g(y) & =  \sum_{\substack{\chrg = m            \\
                            \chrg\ \mathrm{even} \\
                            \text{or}            \\
                            \chrg\ \mathrm{odd}}}^{2n+\eo}
      \tbinom{  \tfrac{\chrg-\eo}2-\ats-\tfrac{m-\eo}2d-u\rl2-\fst}
             {\tfrac{m-\eo}2-\fst} \notag \\
       &      \qquad \qquad {}\times
      \tbinom{n-\tfrac{\chrg-\eo}2-\bts-\tfrac{m-\eo}2d-v\rl2     }
             {\tfrac{m-\eo}2}
                                              y^{\tfrac{-n+\chrg}2-\eo}.
 \end{align}
 Certainly, we may use the line of reasoning that has led us from~\eqref{Eq:GFAn}
 to~\eqref{Eq:GFAnuI}.
 However, we choose another way.
 It is easy to observe that $ n $ and $ \chrg $ form one term
 $ \tfrac{ n + \chrg }2 $ or $ \tfrac{ n - \chrg }2 $
 (see~\eqref{Eq:CdkrI}).
 So we can repeat the steps between~\eqref{Eq:CdkrI} and~\eqref{Eq:gxInt} and obtain
 \begin{equation}
  \label{Eq:gyInt}
  \begin{split}
   g(y) & = (-1)^\alpha y^{ -\tfrac{ n + \eo }2 + a + \tfrac{ m - \eo }2 d }
                ( x - 1 )^{ \alpha + m - \eo + 1 } ( x + 1 )^{ 1 - \fst } \\
        & \qquad
          {}\times
          \frac 1{ 2 \pi i}\oint_\gamma
          \frac{ ( \xi   + 1 )^\fst }
               { ( \xi^2 - 1 )^{ \tfrac{ m - \eo }2 + 1 }
                 ( \xi   - x )^{ \alpha + 1 } } d\xi,
  \end{split}
 \end{equation}
 where we redefine $ \alpha $ and $ a $ as
 \begin{align*}
  \alpha & = n - ( m - \eo ) ( d + 1 ) - \ats - \bts - ( u + v ) \rl2 , \\
   a     & = \ats + u \rl2 .
 \end{align*}
 Note that in this case we do not need the hypergeometric transformation
 that we have performed above.
 Indeed, we already have finite series by definition~\eqref{Eq:gy1}.
 If we replace $ \frac{ m - \eo }2 $ by $ \tilde m $, we get
 \[
  \sum_{ \substack{ m = \eo          \\
                    m\ \mathrm{even} \\
                    \text{or}        \\
                    m\ \mathrm{odd}} }^{ 2 n + \eo } ( \cdot )
  =
  \sum_{ \tilde m = 0 }^n ( \cdot ).
 \]
 After interchanging the order of summation and integration in~\eqref{Eq:Cy1},
 we see that each of two inner sums can be identified with the binomial theorem.
 \begin{equation}
  \label{Eq:Cy2}
  \begin{split}
   & \CCSi_n(y) = (-1)^\alpha y^{ -\tfrac{ n + \eo }2 }
                ( x - 1 )^{ n - \bts + 1 } ( x + 1 )^{ 1 - \ats - \fst } \\
   & \qquad \qquad
                {}\times
                   \frac 1{2\pi i}\oint_\gamma
                    \frac{ ( \xi   + 1 )^\fst }
                         { ( \xi^2 - 1 )
                           ( \xi   - x )^{ n - \ats - \bts + 1 } }       \\
   & \quad
                {}\times
       \left( 1 - \fst
            + \sum_{ \tilde m = 1 }^n
               \left(
                \frac{ \left( \xi   - x \right)^{ 2 ( d + 1) } }
                     { \left(  x^2  - 1 \right)^d
                       \left( \xi^2 - 1 \right)^{ n - \ats - \bts + 1 } }
               \right)^{ \tilde m }
              \vphantom{%
        \left( 1 - \left( \frac{ x - \xi }
                               { x + 1   }
                   \right)^\rl1
        \right)^{ \tilde m - \fst }
        \left( 1 - \left( \frac{ x - \xi }
                               { x - 1   }
                   \right)^\rl1
        \right)^{ \tilde m } }
       \right.                                                            \\
   &
       \left. \vphantom{%
              \sum_{ \tilde m = 1 }^n
               \left(
                \frac{ \left( \xi   - x \right)^{ 2 ( d + 1) } }
                     { \left(  x^2  - 1 \right)^d
                       \left( \xi^2 - 1 \right)^{ n - \ats - \bts + 1 } }
               \right)^{ \tilde m } }
        {}\times%
        \left( 1 - \left( \frac{ x - \xi }
                               { x + 1   }
                   \right)^\rl1
        \right)^{ \tilde m - \fst }
        \left( 1 - \left( \frac{ x - \xi }
                               { x - 1   }
                   \right)^\rl1
        \right)^{ \tilde m }
       \right) d\xi.
  \end{split}
 \end{equation}
 Now, only a finite geometric series remains in~\eqref{Eq:Cy2}.
 Redefine $ h $ as a term of this series
 \begin{equation*}
  \begin{split}
   h & = \frac{ \left( \xi   - x \right)^{ 2 ( d + 1) } }
              { \left(  x^2  - 1 \right)^d
                \left( \xi^2 - 1 \right)^{ n - \ats - \bts + 1 } } \\
     & \qquad \qquad \qquad
        {}\times%
         \left( 1 - \left( \frac{ x - \xi }
                                { x + 1   }
                    \right)^\rl1
         \right)
         \left( 1 - \left( \frac{ x - \xi }
                                { x - 1   }
                    \right)^\rl1
         \right).
  \end{split}
 \end{equation*}
 Substituting $ \frac{ h - h^{ n + 1 } }
                     { 1 - h } $
 for $ \sum_{ \tilde m = 1 }^n h^{ \tilde m } $
 in~\eqref{Eq:Cy2}, we get
 \begin{equation}
  \label{Eq:Cy3}
  \begin{split}
   \CCSi_n(y) & = (-1)^\alpha y^{ -\tfrac{ n + \eo }2 }
                ( x - 1 )^{ n - \bts + 1 } ( x + 1 )^{ 1 - \ats - \fst } \\
              &
        {}\times
           \left(
                   \frac{ 1 - \fst }{2\pi i}\oint_\gamma
                    \frac{ ( \xi   + 1 )^\fst }
                         { ( \xi^2 - 1 )
                           ( \xi   - x )^{ n - \ats - \bts + 1 } } d\xi
           \right.                                                       \\
              & \qquad
           \left. \vphantom{%
                   \frac{ 1 - \fst }{2\pi i}\oint_\gamma
                    \frac{ ( \xi   + 1 )^\fst }
                         { ( \xi^2 - 1 )
                           ( \xi   - x )^{ n - \ats - \bts + 1 } } d\xi }
        {} +       \frac 1{2\pi i}\oint_\gamma
              \Psi( \xi )     d\xi
           -       \frac 1{2\pi i}\oint_\gamma
              \Psi( \xi ) h^n d\xi
           \right),
  \end{split}
 \end{equation}
 where by $ \Psi( \xi ) $ we denote the integrand as follows:
 \begin{align*}
  \Psi( \xi ) &= \frac{ ( \xi   + 1 )^\fst }
                      { ( \xi^2 - 1 )
                        ( \xi   - x )^{ n - \ats - \bts + 1 } } \\
              & \qquad \qquad \qquad {}\times%
     \left( 1 - \left( \frac{ x - \xi }
                            { x + 1   }
                \right)^\rl1
     \right)^{ - \fst }
    \frac{ h }
         { 1 - h }                                              \\
              &= \left( x^2 - 1 \right)^{ k + 1 }
                      ( x   + 1       )^{ \fst \rl2 }           \\
              & \qquad {}\times%
                 \frac{ ( \xi + 1 )^\fst h }
                      { ( \xi - x )^{ n - \ats - \bts + 1 }
                         \psi
                        \left(  ( x +   1 )^{ \rl1 }
                              - ( x - \xi )^{ \rl1 }
                        \right)^\fst },
 \end{align*}
 where
 \begin{equation*}
  \begin{split}
   \psi( \xi )
      & = \left(   x^2 - 1 \right)^{ k + 1 }
          \left( \xi^2 - 1 \right)                               \\
      & \qquad \qquad
          {} - ( x - \xi )^{ 2 ( d + 1 ) }
          \left(   x^2 - 1 \right)^{ \rl1 }                      \\
      & \quad
          {} + ( x - \xi )^{ ( d + 1 ) + ( k + 2 ) }
          \left( ( x - 1 )^{ \rl1 } + ( x + 1 )^{ \rl1 } \right) \\
      & \qquad \qquad \qquad \qquad \qquad \qquad
          {} - ( x - \xi )^{ 2 ( k + 2 ) }
  \end{split}
 \end{equation*}
 such that $ 1 - h = \frac{ \psi( \xi ) }
                          { \left(   x^2 - 1 \right)^{ k + 1 }
                            \left( \xi^2 - 1 \right) } $.
 Now if we recall that the contour $ \gamma $ was initially considered
 as a small circle surrounding the point $ x $,
 then, using the residue theorem, we get
 \begin{equation}
  \label{Eq:Res2y}
  \begin{split}
   \frac 1{2\pi i}\oint_\gamma &
    \Psi( \xi ) d\xi
     =            \left( x^2 - 1 \right)^{ k + 1 }
                       ( x   + 1       )^{ \fst \rl2 }           \\
                               &
   {}\times%
   \sum_{ j = 1 }^{ 2 ( k + 2 ) }
    \frac{ \displaystyle
            \lim_{ \xi \to \xi_j } h }
         { \left( \xi_j - x \right)^{ n - \ats - \bts + 1 }
           \displaystyle
            \prod_{\substack{ m = 1 ,\\ m \neq j } }^{ 2 ( k + 2 ) }
             \left( \xi_j - \xi_m \right) }                      \\
                               &
    {}\times
    \left(
     \begin{cases}
      \dfrac{ \xi_j + 1 }
            {   ( x +   1   )^{ \rl1 }
              - ( x - \xi_j )^{ \rl1 } } ,     & \xi_j \neq -1 , \\
      \dfrac 1
            { \rl2
                ( x - \xi_j )^{ \rl1 - 1 } } , & \text{otherwise},
     \end{cases}
    \right)^\fst,
  \end{split}
 \end{equation}
 where
 \[
  \lim_{ \xi \to \xi_j } h
   =
    \begin{cases}
     \dfrac{ \rl1
                      \left( x - \xi_j \right)^{ d + 1 }
             \left(   \left( x - \xi_j \right)^{ \rl1 }
                    - \left( x + \xi_j \right)^{ \rl1 }
             \right) }
           {  2 \xi_j \left( x + \xi_j \right)^{ k + 1 } } , & | \xi_j | = 1 , \\
     1 ,                                                     & \text{else},
    \end{cases}
 \]
 and $ \xi_1 $, $ \xi_2 $, $ \dots $, $ \xi_{ 2 ( k + 2 ) } $ are roots of
 $ \psi( \xi ) $.

 It can easily be checked that
 \begin{equation}
  \label{Eq:Res3y}
  \frac 1{2\pi i}\oint_\gamma
   \Psi( \xi ) h^n d\xi = 0.
 \end{equation}
 We might prove it using the contour integration methods.
 But more simply is to observe that $ \CCS_n^\chrg = 0 $ whenever $ \chrg > n $.

 It remains to find the first integral in~\eqref{Eq:Cy3}. Obviously,
 \begin{equation}
  \label{Eq:Res1y}
  \begin{split}
   & \frac{ 1 - \fst }{2\pi i}\oint_\gamma
      \frac{ ( \xi   + 1 )^\fst }
           { ( \xi^2 - 1 )
             ( \xi   - x )^{ n - \ats - \bts + 1 } } d\xi \\
   &
     = ( 1 - \fst )
        \left(
                 \frac 1 { 2 ( -1 - x )^{ n - \ats - \bts + 1 } }
               - \frac 1 { 2 (  1 - x )^{ n - \ats - \bts + 1 } }
        \right).
  \end{split}
 \end{equation}
 Note that these integrals become equal to $ 0 $
 whenever $ n - \ats - \bts + 1 \leq 0 $.

 Substituting~\eqref{Eq:Res1y}, \eqref{Eq:Res2y}, and~\eqref{Eq:Res3y}
 for the first, for the second, and for the third integrals in~\eqref{Eq:Cy3},
 we obtain the final result.

 \section{Algorithms for encoding and decoding
  constant-weight and constant-charge
  binary run-length limited sequences}
  \label{Sec:EnumAlg}

 Recall that $\Sdklr$ denotes a set of constant-weight or constant-charge
 binary run-length constrained sequences
 $\xn=(x_1, x_2, \dots,$ $ x_n)$ of length $n$.
 Let the set $\Sdklr$ be ordered lexicographically.
 From~\cite{Cover73} it follows that the lexicographic index of
 $\xn \in \Sdklr$ is given by
 \begin{equation}
  \label{Eq:Cover1}
  N(\xn)=\sum_{j=1}^n x_j W( \pn ) ,
 \end{equation}
 where $W( \pn )$ denotes the number of sequences
 in $\Sdklr$ with given prefix $ \pn = (x_1, x_2, \dots, x_{j-1}, 0) $.

 The decoding algorithm, for given sequence $\xn$,
 find its lexicographic index $N$, $0 \le N < |\Sdklr|$.
 This is done by successive approximation method~\cite{Cover73, BImmink83}
 using~\eqref{Eq:Cover1} and $W( \pn )$ as the number of
 sequences in $\Sdklr$ with given prefix $ \pn = (x_1, x_2, \dots,$ $x_{j-1}, 0) $.

 By $ a_j ( \pn ) $ denote the number of trailing zeros of the prefix \pn.
 By $ \nu_j   ( \pn ) = \nu_{ j - 1 } = \sum_{ i = 1 }^j x_i $ and
 by $ \chrg_j ( \pn ) =                 \sum_{ i = 1 }^j ( -1 )^{ \nu_i } $
 denote the weight and the charge of this prefix.
 Since $ \pn $ is the prefix of $ \xn $,
 then subsequence $ \tilde \xn = (x_j, x_{j+1}, \dots, x_n) $
 is the rest of $ \xn $,
 and $ l_j $ is the leading run of zeros in this subsequence.
 We define $ l_j $ as the complement of $ a_j ( \pn ) $
 in $ l $ (for leading run of zeros) or
 in $ k $ (for another run of zeros) as follows:
 \[
  l_j=\begin{cases}
   l-a_j(\pn), & \nu_{ j - 1 } = 0 , \\
   k-a_j(\pn), & \text{otherwise} .
  \end{cases}
 \]
 Similarly, by $ r_j $ we denote the trailing run of zeros
 in the subsequence $ \tilde \xn $.
 If this subsequence consists of zeros, this may mean either
 the leading run of zeros (when $ \nu = 0 $), or the trailing run of zeros.
 Therefore, we define $ r_j $ as the complement of $ a_j ( \pn ) $ either
 in $ \min( l , r ) $, or in $ r $.
 In the case of nonzero $ \tilde \xn $, we define $ r_j $ as $ r $, i.e.,
 \[
  r_j=\begin{cases}
   \min(l, r)-a_j(\pn), & \nu = \nu_{ j - 1 }  =   0 , \\
   r-a_j(\pn),          & \nu = \nu_{ j - 1 } \neq 0 , \\
   r,                   & \nu \neq \nu_{ j - 1 } ,
  \end{cases}
 \]
 or
 \[
  r_j=\begin{cases}
   \min(l, r)-a_j(\pn), & \chrg_{n-j}(\pn)=n-j \text{ and } \nu_{j-1}=0,\\
   r-a_j(\pn),          & \chrg_{n-j}(\pn)=n-j \text{ and } \nu_{j-1} \neq 0,\\
   r,                   & \chrg_{n-j}(\pn) \neq n-j ,
  \end{cases}
 \]
 where
 \[
  \chrg_{n-j}(\pn)=(-1)^{\nu_{j-1}}(\chrg-\chrg_{j-1})-1.
 \]
 Then we can compute the number of constant-weight sequences $ W_\nu ( \pn ) $ as
 \[
  W_\nu ( \pn )
  = \begin{cases}
     \CWSl_{n-j}^{\nu-\nu_j}(d, k, l_j, r_j), & l_j \geq 0 \text{ and }%
                                                r_j \geq 0 ,    \\
     0,                                       & \text{otherwise}
    \end{cases}
 \]
 and the number of constant-charge sequences $ W_\chrg ( \pn ) $ as
 \[
  W_\chrg ( \pn )
  = \begin{cases}
     \CCSl_{n-j}^{\chrg_{n-j}(\pn)}(d, k, l_j, r_j), & l_j \geq 0 \text{ and }%
                                                       r_j \geq 0 ,    \\
     0,                                              & \text{otherwise}.
    \end{cases}
 \]

 Let $\boldsymbol{\gamma}=(\gamma_0, \gamma_1, \dots, \gamma_n)$
 be a binary vector; components of this vector
 $\gamma_j$, $0 \leq j \leq n$ indicate that $ \xn $
 does     (if $\gamma_j=1$) or
 does not (if $\gamma_j=0$)
 have the weight $ \nu = j $ or
      the charge $ \chrg = 2 j - n $.

 Hence, the number $W(\pn)$ of sequences with given prefix \pn\ be
 \begin{equation*}
  W(\pn) = \begin{cases}
            \displaystyle
            \sum_{\nu=0}^n
             \gamma_\nu
              W_\nu (\pn)  , & \parbox[t]{.27\linewidth}{ constant-weight case, } \\
            \displaystyle
            \sum_{ \substack{ \chrg = -n            \\
                              \chrg \ \mathrm{even} \\
                              \text{or}             \\
                              \chrg \ \mathrm{odd} } }^{ n }
             \gamma_{ ( \chrg + n ) / 2 }
              W_\chrg (\pn), & \parbox[t]{.27\linewidth}{ constant-charge case. }
           \end{cases}
 \end{equation*}

 We introduce the next variables: $a$, $w$, $ c $,
 which correspond to $ a_j  ( \pn ) $, $ \nu_{ j - 1 } $, $ \chrg_{ j - 1 } $.

 \begin{tabbing}
  \qquad \= \quad \= \quad \= \+ \kill
  $ N:=0; \quad a:=1; \quad w:=0; \quad c := 0 ; $ \\
  \textbf{for} $j:=1$ \textbf{to} $n$ \textbf{do} \+ \\
   \pushtabs
   Get $W( \pn );$\\
   \poptabs
   \textbf{if} $ x_j = 1 $ \textbf{then}\\
    \> $N:=N+W( \pn );$\\
    \> $a:=1; \quad w:=w+1;$\\
   \textbf{else}\\
    \> $a:=a+1;$\\
   \textbf{end if \dots else} \\
   $ c := c + ( -1 )^w ; $ \- \\
 \textbf{end for}.
 \end{tabbing}

 The encoding (inverse) algorithm, for given lexicographic index $N$,
 $0 \le N < |\Sdklr|$, find the corresponding $\xn$.
 \begin{tabbing}
  \qquad \= \quad \= \quad \= \+ \kill
  $ a:=1; \quad w:=0; \quad c := 0 ; $ \\*
  \textbf{for} $j:=1$ \textbf{to} $n$ \textbf{do} \+ \\*
   \pushtabs
   Get $W( \pn );$\\*
   \poptabs
   \textbf{if} $ N \geq W( \pn ) $ \textbf{then}\\*
    \> $N:=N-W( \pn );$\\*
    \> $x_j:=1; \quad a:=1; \quad w:=w+1;$\\*
   \textbf{else}\\*
    \> $x_j:=0; \quad a:=a+1;$\\*
   \textbf{end if \dots else}\\*
   $ c := c + ( -1 )^w ; $ \- \\*
  \textbf{end for}.
 \end{tabbing}

 \section{Further remarks}
  \label{Sec:FuRem}

  \subsection{}
 As it was shown in Section~\ref{Sec:Recur},
 the recursion relation~\eqref{Eq:Cc}
 has an implicit mutual nature.
 We can cite an example of practically identical coding scheme~\cite{Kurmaev002}
 in which we obtained mutual recursions in explicit form.
 Alternating runs in~\cite{Kurmaev002} were presented
 by series of zeros or ones with independent constraints.
 It seems now; there was a heavy construction without significant preference
 of present constant-charge code.

  \subsection{}
 Channels may cause peak shifts in $dk$-constrained sequences~\cite{Shamai91}.
 It is considered as a more frequent error.
 We provide an example (see Table~\ref{Tab:PeakShift}) which shows
 that the charge distribution seems rather suitable
 for the peak shift analysis then the weight distribution.
 On the other hand, the weight distribution remains useful
 for erasure and insertion control.
 In more detail, consider the integer
 (or composition) representation of RLL sequences~\cite{Zehavi88}.
 This representation is used in error detection and correction technique
 including the peak shifts correction~\cite{Ferreira91, Levenshtein93}.
 In such case an RLL sequence is parsed uniquely
 into a concatenation of phrases, each phrase beginning with one.
 By $ \phi_j $ denote the length of $ j $th phrase.
 We can state a simple relation between $ \phi_j $ and charge $ \chrg_j $ as follows:
 \[
  \phi_j = j - i
         = \frac{ \chrg_j - \chrg_i }
                { ( -1 )^{ \nu_j } },
 \]
 where $ i $ and $ j $ are positions of consecutive ones such that $ i < j $.
 Moreover, from Section~\ref{Sec:Recur}
 and Section~\ref{Sec:EnumAlg} it follows that Cover's enumerative scheme
 implies counting of $ \phi_j $.

 \begin{table}[!t]
  \renewcommand{\arraystretch}{1.3}
  \caption{An Example of Single Peak Shift}
  \label{Tab:PeakShift}
  \centering
  \renewcommand{\arraystretch}{0}
  \newcommand{\z}{$\scriptstyle1$}
  \newcommand{\xs}{\rule{0pt}{2ex}}
  \newcommand{\zs}{\rule{0pt}{2ex}}
  \newcommand{\zf}{\rule{0pt}{2pt}}
  \newcommand{\bo}{$\boldsymbol{1}$}
  \tabcolsep=0.25em
   \begin{tabular}{|c|cccccccc|lr|c|}
    \hline
    \strut %
    $N$ & $x_1$ & $x_2$ & $x_3$ & $x_4$ & %
          $x_5$ & $x_6$ & $x_7$ & $x_8$ & $\nu$ & $\chrg$ &\\
    \hhline{|=|========|==|=|}
    \xs%
    0  &  0&  1&  0&  0&  0&  0&\bo&  0 &   2   &         & right    \\
    \zs%
       & \z&-\z&-\z&-\z&-\z&-\z& \z& \z &       &  -2     & shift    \\
    \zf&   &   &   &   &   &   &   &    &       &         &          \\
    \hline
    \xs%
    1  &  0&  1&  0&  0&  0&\bo&  0&  0 &   2   &         & true     \\
    \zs%
       & \z&-\z&-\z&-\z&-\z& \z& \z& \z &       &   0     & position \\
    \zf&   &   &   &   &   &   &   &    &       &         &          \\
    \hline
    \xs%
    2  &  0&  1&  0&  0&\bo&  0&  0&  0 &   2   &         & left     \\
    \zs%
       & \z&-\z&-\z&-\z& \z& \z& \z& \z &       &   2     & shift    \\
    \zf&   &   &   &   &   &   &   &    &       &         &          \\
    \hline
   \end{tabular}
 \end{table}

  \subsection{}
 Since run-length constraints bound the weight distribution,
 we have $\nu \in [\nu_{\min}, \nu_{\max}]$,
 where          $0 \le \nu_{\min} \le \nu_{\max} \le n$.
 In~\cite{Ytrehus91}, Ytrehus obtained
 \[
  \nu_{\min}=\begin{cases}
   0,                         & n \le \min(l, r),\\
   1,                         &       \min(l, r)<n \le l+r+1,\\
   \left\lceil \dfrac{n-l-r-1}{k+1} \right\rceil +1, & l+r+1<n
  \end{cases}
 \]
 and
 \[
  \nu_{\max}=\left\lceil \frac{n}{d+1} \right\rceil
 \]
 for $ dklr $ sequences.
 Similarly, for the charge distribution
 of $ dk $ sequences we can write
 \[
  |\chrg|_{\max} = \begin{cases}
                    n - 2 m ( d + 1 ) ,         & \begin{aligned}[t]
                                                   0     & \leq n - m ( k + d + 2 ) \\
                                                         & \leq k + 1 ,
                                                  \end{aligned} \\
                    2 ( m + 1 ) \\ \quad {}\times%
                                ( k + 1 ) - n , & \begin{aligned}[t]
                                                   k + 1 & \leq n - m ( k + d + 2 ) \\
                                                         & \leq k + d + 2 ,
                                                  \end{aligned}
                   \end{cases}
 \]
 where
 \[
  m = \left\lfloor \frac{ n }{ k + d + 2 } \right\rfloor .
 \]

  \subsection{}
 The charge of the prefix $ \chrg_j $
 (see Section~\ref{Sec:Recur} and Section~\ref{Sec:EnumAlg})
 is called the running digital sum (RDS),
 see~\cite{Pierobon84}.
 The RDS has a finite range of values.
 For the RLL sequences with zero accumulated charge ($ \chrg_n = 0 $),
 we can see that absolute value of RDS does not exceed
 $ \left\lfloor |\chrg|_{\max} / 2 \right\rfloor + 1 $.

 We can impose restrictions on the range of RDS values.
 Denote by $ \BRDS_1 $ and $ \BRDS_2 $ the lower and the upper bounds
 of this constrained range.
 The range of RDS bounded by $ \BRDS_1 $ and $ \BRDS_2 $
 is said to be the digital sum variation (DSV),
 see~\cite{Immink00, Braun00anenumerative}.
 After the authors~\cite{Braun00anenumerative},
 DSV constrained RLL sequences are called DCRLL sequences.

 By $\CCS_n^\chrg (d, k, r , \BRDS_1 , \BRDS_2 )$
 denote the number of DCRLL sequences beginning with one.
 Also by $\CCSl_n^\chrg (d, k, l, r , \BRDS_1 , \BRDS_2 )$ we
 denote the number of DCRLL sequences beginning with a leading run of zeros.
 Below under $\CCS_n^\chrg ( \BRDS_1 , \BRDS_2 )$
 we consider $\CCS_n^\chrg (d, k, r , \BRDS_1 , \BRDS_2 )$
 and under $\CCSl_n^\chrg$
 we similarly consider $\CCSl_n^\chrg (d, k, l, r , \BRDS_1 , \BRDS_2 )$.

 As shown in~\cite{Vasilev91en}, calculation of the number of DCRLL sequences
 can be performed as concatenation of two subcodes;
 the volume of each subcode is found recurrently in explicit mutual form.

 The recurrent method for calculating the number of constant-charge RLL sequences,
 which we suggest in Section~\ref{Sec:Recur},
 allows us simply turn to calculating the number of DCRLL sequences.
 Indeed, we know the charge value of the prefix $ \pn $
 at the each level of recursion~\eqref{Eq:Cc}.
 Therefore we can control DSV by using the additional condition
 $ \BRDS_{ 1 m } \leq \chrg_m \leq \BRDS_{ 2 m } $.
 For this reason, we do not account $ \CCS_m^{ \chrg_m } $
 for which this condition does not satisfy.
 Here under  $ \BRDS_{ 1 m } $ and $ \BRDS_{ 2 m } $
 we consider $ \BRDS_1 $       and $ \BRDS_2 $
 after justification at the each level of recursion~\eqref{Eq:Cc}.
 Moreover, each level of this recursion
 alternates the direction of charge changing;
 then it is sufficient to check either $ \BRDS_{ 1 m } \leq \chrg_m $
                                    or $ \chrg_m \leq \BRDS_{ 2 m } $ condition
 depending on the direction of charge changing.
 By substituting $ \BRDS_1 $ for $ \BRDS_2 $ and vice versa,
 when calling $ \CCS_n^\chrg ( \BRDS_1 , \BRDS_2 ) $,
 we keep the only condition $ \BRDS_{ 1 m } \leq \chrg_m $
 and obtain the implicit mutual recursion similar to~\eqref{Eq:Cc}.

 From initial conditions~\eqref{Eq:Cmn}
 it follows that a unique sequence of zero length and zero charge exists.
 In turn, from this statement it follows that we need an additional condition;
 this condition allows us to take into account the existence of
 $ \CCS_0^0 ( \BRDS_1 , \BRDS_2 ) $.
 Indeed, the sequences, which beginning with zero,
 have initial charge equal to $ 1 $,
 the sequences, which beginning with one,
 have initial charge equal to $ -1 $,
 and the sequence of zero length,
 have initial charge equal to $ 0 $.
 We now write this triple condition using the Iverson bracket notation
 \[
  [  \text{condition}  ] = \begin{cases}
                            1 , & \text{the condition is true} , \\
                            0 , & \text{the condition is false}.
                           \end{cases}
 \]

 Thus we can rewrite Proposition~\ref{Prop:C} as
 \begin{proposition}
  The numbers $\CCS_n^\chrg ( \BRDS_1 , \BRDS_2 )$ and $\CCSl_n^\chrg$
  can be obtained as:\\
  If $ -[ n \neq 0 ] > \BRDS_2 $, then
  \[
   \CCS_n^\chrg ( \BRDS_1 , \BRDS_2 ) = 0 .
  \]
  If $ \chrg     = -n $ and the sequences begin with one, then
  \begin{equation*}
   \CCS_n^\chrg ( \BRDS_1 , \BRDS_2 ) =
    \begin{cases}
     1 , & n \leq \min( r+1 , -\BRDS_1 ) , \\
     0 , & \text{otherwise} .
    \end{cases}
  \end{equation*}
  If $ \chrg \neq  -n $ and the sequences begin with one, then
  \begin{equation*}
   \begin{split}
    & \CCS_n^\chrg ( \BRDS_1 , \BRDS_2 ) \\
    & \; = %
    \begin{cases}
     \displaystyle{%
     \sum_{ j = d + 1 }^{ \min( n , k + 1 , -\BRDS_1 ) } }
      \CCS_{ n - j }^{ -\chrg - j } ( -\BRDS_2 - j , -\BRDS_1 - j ) , & d + 1 \leq n , \\
     0 ,                                                              & \text{otherwise} .
    \end{cases}
   \end{split}
  \end{equation*}
  If $ \BRDS_1 > [ n \neq 0 ] $, then
  \[
   \CCSl_n^\chrg = 0 .
  \]
  If $ \chrg     =  n $ and a leading series is running, then
  \[
   \CCSl_n^\chrg =
    \begin{cases}
     1 , & n \leq \min( l , r , \BRDS_2 ) , \\
     0 , & \text{otherwise}.
    \end{cases}
  \]
  If $\chrg \neq    n$ and a leading series is running, then
  \begin{equation*}
   \CCSl_n^\chrg = \CCS_n^\chrg ( \BRDS_1 , \BRDS_2 ) +
     \sum_{ j = 1 }^{ \min( n , l , \BRDS_2 ) }
      \CCS_{ n - j }^{ \chrg - j } ( \BRDS_1 - j , \BRDS_2 - j ) .
  \end{equation*}
 \end{proposition}
 Evidently, the algorithms from Section~\ref{Sec:EnumAlg}
 are suitable for encoding and decoding DCRLL sequences
 if $ W_\chrg ( \pn ) $ will be calculated as
 \[
  \begin{split}
   & W_\chrg ( \pn ) \\
   & \; = %
    \begin{cases}
     \CCSl_{n-j}^{\chrg_{n-j}(\pn)}(d, k, l_j, r_j,
                  \BRDS_{ 1 j } , \BRDS_{ 2 j }   ), & \begin{gathered}[t]
                                                        l_j \geq 0 \text{ and }%
                                                        r_j \geq 0 ,       \\
                                                        \text{%
                                                         \textnormal{and}} \\[-.5ex]
                                                        \BRDS_1 \leq \chrg_j ( \pn ) %
                                                                \leq \BRDS_2 ,
                                                       \end{gathered} \\[5ex]
     0,                                              & \text{otherwise}.
    \end{cases}
   \end{split}
 \]
 where
 \begin{align*}
  \chrg_j ( \pn ) & = \chrg_{j-1} + (-1)^{\nu_{j-1}} ,   \\
  \BRDS_{ 1 j }   & = \left( \begin{cases}
                                \BRDS_1 , & \text{if $ \nu_{j-1} $ is even} , \\
                              - \BRDS_2 , & \text{otherwise} ;
                             \end{cases}
                      \right)
                      - (-1)^{\nu_{j-1}} \chrg_j (\pn) , \\
  \BRDS_{ 2 j }   & = \frac{ (-1)^{\nu_{j-1}} \left( \BRDS_1 + \BRDS_2 \right)
                                                   - \BRDS_1 + \BRDS_2 } 2
                      - (-1)^{\nu_{j-1}} \chrg_j (\pn) .
 \end{align*}

  \subsection{}
 One can find in literature
 some examples of application of generating functions for RLL coding.
 From 48 years range of publications~\cite{Gilbert60} -- \cite{Choi08},
 we cite two examples.
 Kolesnik and Krachkovsky~\cite{Kolesnik91},
 when deriving the Gilbert-Varshamov bound,
 estimated the volume $ V_r( \xn ) $ of sphere in $ \Sdkr(n) $
 centered on $ \xn $ as
 \[
  V_r( \xn ) \leq \min_{ 0 \leq y \leq 1 } \frac{     \CWS_n( y ) }
                                                { y^r \CWS_n( 0 ) },
 \]
 where $ \CWS_n( y ) $ may be taken from~\eqref{Eq:GFAnu}.
 Note, that practical applications of the generating functions
 for enumerating RLL sequences often require a rational expression.
 Ferreira and Lin~\cite{Ferreira91}
 and some other authors~\cite{Immink97, Kurmaev02en}
 obtained a generating function for enumerating $ dk $ and $ dkr $ sequences
 \[
     \Sdkr(t) =   \frac{ t(1-t^{r+1})}
                       {1-t- t^{d+1}+ t^{k+2}}.
 \]
 Assume $ y = 1 $. Then from~\eqref{Eq:GFA} we immediately get the same.

  \subsection{}
 Recall~\eqref{Eq:ResIu}, \eqref{Eq:R}, and~\eqref{Eq:ResIv}.
 One can identify $ I_u $ and $ I_v $
 as generating functions of the Jacobi polynomials~\cite{abramowitz+stegun}.
 Then $ \cos\left(\uvcd \varphi\right) $
 and  $ \sin\left(\uvcd \varphi\right) $
 in~\eqref{Eq:J14ImFin} and in~\eqref{Eq:J23ImFin}
 can be identified with Chebyshev polynomials 
 in the variable $ \frac{P_0}
                        {2\sqrt{\bd\bk}(\xi^2-1)} $.
 The technique which uses the theory of orthogonal polynomials
 is known in charge constrained coding~\cite{KerpezGH92}.

  \subsection{}
 We obtain the generating function~\eqref{Eq:Ct1}
 in the form of elliptic integrals~\eqref{Eq:J14ImFin} and~\eqref{Eq:J23ImFin}.
 Further exploration may require a canonical form of these integrals.
 Using factorization, it is not so hard to reduce
 the first term of the integrands in~\eqref{Eq:J14ImFin} and~\eqref{Eq:J23ImFin}
 to rational functions.
 In such case, we shall use polynomial expansion
 of $ \cos\left(\uvcd \varphi\right) $.
 The next transformations, which bring these integrals to canonical form,
 may be taken from~\cite{abramowitz+stegun}.

  \subsection{}
 If we need a two-variable generating function
 for the case of constant-charge sequences
 \[
  \CCSi( t , y) = \sum_{\substack{ \chrg = \cl              \\
                                   \chrg \ \mathrm{even \ } \\
                                          \text{or}         \\
                                   \chrg \ \mathrm{odd \ } } }^{ \cu }
                    \sum_{\substack{ n = \eo           \\
                                     n \ \mathrm{even} \\
                                        \text{or}      \\
                                     n \ \mathrm{odd}}}^\infty
                           \CCSi_n^\chrg t^{ ( n - \eo ) / 2 } y^{ ( \chrg - \eo ) / 2 } ,
 \]
 then we can write
 \[
  \CCSi( t , y ) = \sum_{\substack{ \chrg = \cl           \\
                                    \chrg \ \mathrm{even} \\
                                           \text{or}      \\
                                    \chrg \ \mathrm{odd} } }^{ \cu }
                  \CCSi^\chrg( t ) y^{ ( \chrg - \eo ) / 2 } .
 \]
 In such case we may perform summation of~%
 \eqref{Eq:IntegrandJ1}, \eqref{Eq:IntegrandJ2}, \eqref{Eq:IntegrandJ3},
 and~\eqref{Eq:IntegrandJ4}
 over the range $ [ \cl , \cu ] $ of even or odd numbers.
 In result we do not obtain any additional poles with nonzero residues.
 So, there remain the contours of integration
 depicted on Fig~\ref{Fig:DumbbellContour}.

 \section{Conclusion}
  \label{Sec:Concl}

 We have presented constant-weight and constant-charge
 run-length constrained binary sequences.
 On the base of Cover's enumerative technique,
 we have obtained recursion relations for calculating
 the numbers of these sequences.
 For investigation of the asymptotic behavior of these values,
 we have derived generating functions for enumerating such sequences.
 We have proved that generating function for enumerating
 constant-charge sequences does not be expressed in a closed form.
 So, we have presented the long chain of derivation steps
 which led us to expression for this generating function
 in the form of elliptic integrals.
 Also we have described two algorithms
 for enumerative encoding and decoding
 constant-weight and constant-charge sequences.
 Then, we have provided some examples of application for our results;
 in particular, we have extended our results on RDS constrained sequences.

 \bibliographystyle{IEEEtran}
 \bibliography{rll-seq}

\begin{thebibliography}{10}
\providecommand{\url}[1]{#1}
\csname url@samestyle\endcsname
\providecommand{\newblock}{\relax}
\providecommand{\bibinfo}[2]{#2}
\providecommand{\BIBentrySTDinterwordspacing}{\spaceskip=0pt\relax}
\providecommand{\BIBentryALTinterwordstretchfactor}{4}
\providecommand{\BIBentryALTinterwordspacing}{\spaceskip=\fontdimen2\font plus
\BIBentryALTinterwordstretchfactor\fontdimen3\font minus
  \fontdimen4\font\relax}
\providecommand{\BIBforeignlanguage}[2]{{%
\expandafter\ifx\csname l@#1\endcsname\relax
\typeout{** WARNING: IEEEtran.bst: No hyphenation pattern has been}%
\typeout{** loaded for the language `#1'. Using the pattern for}%
\typeout{** the default language instead.}%
\else
\language=\csname l@#1\endcsname
\fi
#2}}
\providecommand{\BIBdecl}{\relax}
\BIBdecl

\bibitem{Immink2004}
K.~A.~S. Immink, \emph{\BIBforeignlanguage{english}{Codes for Mass Data Storage
  Systems}}, 2nd~ed.\hskip 1em plus 0.5em minus 0.4em\relax Eindhoven, The
  Netherlands: Shannon Foundation Publishers, 2004.

\bibitem{ImminkSiegelWolf98}
K.~A.~S. Immink, P.~H. Siegel, and J.~K. Wolf, ``Codes for digital recorders,''
  \emph{{IEEE} Trans. Inf. Theory}, vol.~44, no.~6, pp. 2260--2299, Oct 1998.

\bibitem{Cover73}
T.~M. Cover, ``Enumerative source coding,'' \emph{{IEEE} Trans. Inf. Theory},
  vol. IT-19, no.~1, pp. 73--77, Jan. 1973.

\bibitem{Immink97}
K.~A.~S. Immink, ``A practical method for approaching the channel capacity of
  constrained channels,'' \emph{{IEEE} Trans. Inf. Theory}, vol. IT-43, no.~5,
  pp. 1389--1399, Sep. 1997.

\bibitem{Schalkwijk72}
J.~P.~M. Schalkwijk, ``An algorithm for source coding,'' \emph{{IEEE} Trans.
  Inf. Theory}, vol. IT-18, no.~3, pp. 395--399, May 1972.

\bibitem{Ytrehus91}
{\O}.~Ytrehus, ``Upper bounds on error-correcting runlength-limited block
  codes,'' \emph{{IEEE} Trans. Inf. Theory}, vol. IT-37, no.~3, pp. 941--945,
  May 1991.

\bibitem{Kurmaev02en}
O.~F. Kurmaev, ``Enumerative coding for constant-weight binary sequences with
  constrained run-length of zeros,'' \emph{Problems of Information
  Transmission}, vol.~38, no.~4, pp. 249--254, 2002.

\bibitem{Lee88}
P.~Lee, ``Combined error-correcting/modulation recording codes,'' Dr.\ scient.\
  thesis, University of California, San Diego, Mar. 1988.

\bibitem{Forsberg88}
K.~Forsberg and I.~Blake, ``The enumeration of (d,k) sequences,'' in
  \emph{Proc. 26th Allerton Conf. on Communications, Control, and Computing},
  Montecello, IL., Sep.~28--30 1988, pp. 471--472.

\bibitem{Huffman52}
D.~A. Huffman, ``A method for the construction of minimum-redundancy codes,''
  \emph{Proc. IRE}, vol.~40, pp. 1098--1101, Sep. 1952.

\bibitem{Riordan58}
J.~Riordan, \emph{An introduction to combinational analysis}.\hskip 1em plus
  0.5em minus 0.4em\relax New-York: Wiley, 1958.

\bibitem{Kolesnik91}
V.~D. Kolesnik and V.~Y. Krachkovsky, ``Generating functions and lower bounds
  on rates for limited error-correcting codes,'' \emph{{IEEE} Trans. Inf.
  Theory}, vol. IT-37, no.~3, pp. 778--788, May 1991.

\bibitem{Szego75}
G.~Szeg{\"o}, \emph{Orthogonal Polynomials}, 4th~ed.\hskip 1em plus 0.5em minus
  0.4em\relax Providence, RI: Amer.\ Math.\ Soc.\ Colloq.\ Publ., 1975.

\bibitem{abramowitz+stegun}
M.~{Abramowitz} and I.~A. {Stegun}, \emph{Handbook of Mathematical Functions
  with Formulas, Graphs, and Mathematical Tables}, 9th~ed.\hskip 1em plus 0.5em
  minus 0.4em\relax New York: Dover, 1964.

\bibitem{BImmink83}
G.~F.~M. Beenker and K.~A.~S. Immink, ``A generalized method for encoding and
  decoding run-length-limited binary sequences,'' \emph{{IEEE} Trans. Inf.
  Theory}, vol. IT-29, no.~3, pp. 751--754, May 1983.

\bibitem{Kurmaev002}
O.~Kurmaev, ``An enumerative method for encoding and decoding constant-weight
  run-length limited binary sequences,'' in \emph{IEEE Intl. Symposium on
  Information Theory}, Lausanne, Switzerland, Jun.~6~-- Jul.~5 2002, p. 328.

\bibitem{Shamai91}
S.~Shamai and E.~Zehavi, ``Bounds on the capacity of the bit-shift magnetic
  recording channel,'' \emph{{IEEE} Trans. Inf. Theory}, vol. IT-37, no.~3, pp.
  863--872, May 1991.

\bibitem{Zehavi88}
E.~Zehavi and J.~K. Wolf, ``On runlength codes,'' \emph{{IEEE} Trans. Inf.
  Theory}, vol. IT-34, no.~1, pp. 45--54, Jan. 1988.

\bibitem{Ferreira91}
H.~C. Ferreira and S.~Lin, ``Error and erasure control $(d, k)$ block codes,''
  \emph{{IEEE} Trans. Inf. Theory}, vol. IT-37, no.~5, pp. 1399--1408, Sep.
  1991.

\bibitem{Levenshtein93}
V.~I. Levenshtein and A.~J.~H. Vink, ``Perfect $(d, k)$-codes capable of
  correcting single peak-shifts,'' \emph{{IEEE} Trans. Inf. Theory}, vol.
  IT-39, no.~2, pp. 656--662, Mar. 1993.

\bibitem{Pierobon84}
G.~L. Pierobon, ``Codes for zero spectral density at zero frequency,''
  \emph{{IEEE} Trans. Inf. Theory}, vol. IT-30, no.~2, pp. 435--439, Mar. 1984.

\bibitem{Immink00}
K.~A.~S. Immink, ``{DC}-free codes of rate $ ( n - 1 ) / n $, $ n $ odd,''
  \emph{{IEEE} Trans. Inf. Theory}, vol. IT-46, no.~2, pp. 633--634, Mar. 2000.

\bibitem{Braun00anenumerative}
V.~Braun and K.~A.~S. Immink, ``An enumerative coding technique for {DC}-free
  runlength-limited sequences,'' \emph{{IEEE} Trans. Commun.}, vol.~48, no.~12,
  pp. 2024--2031, Dec. 2000.

\bibitem{Vasilev91en}
P.~I. Vasil'ev, ``Block run-length-limited coding for digital magnetic
  recording,'' Ph.D. dissertation, LIAP, Leningrad, USSR, 1991, (in Russian).

\bibitem{Gilbert60}
E.~Gilbert, ``Synchronization of binary messages,'' \emph{IRE Trans.\ Inform. \
  Theory}, vol.~6, no.~4, pp. 470--477, Sep. 1960.

\bibitem{Choi08}
Y.~Choi and W.~Szpankowski, ``Pattern matching in constrained sequences,'' in
  \emph{IEEE Intl. Symposium on Information Theory}, Toronto, Canada,
  Jul.~6--11 2008, pp. 2141--2145.

\bibitem{KerpezGH92}
K.~J. Kerpez, A.~Gallopoulos, and C.~Heegard, ``Maximum entropy
  charge-constrained run-length codes,'' \emph{{IEEE} J. Sel. Areas Commun.},
  vol. SAC-10, no.~1, pp. 242--253, Jan. 1992.

\end{thebibliography}
\end{document}